\documentclass[useAMS,usenatbib]{mn2e}

\def\equ#1{eq.~(\ref{eq:#1})}
\def\se#1{\S\ref{sec:#1}}
\def\Fig#1{Fig.~\ref{fig:#1}}
\def\etal{{\it et al.\ }}

\def\la{\langle}
\def\ra{\rangle}
\def\be{\begin{equation}}
\def\ee{\end{equation}}
\def\prop{\propto}
\def\ifm#1{\relax\ifmmode#1\else$\mathsurround=0pt #1$\fi}
\def\kms{\ifmmode\,{\rm km}\,{\rm s}^{-1}\else km$\,$s$^{-1}$\fi}

\def\d{{\rm d}}
\def\dd{{\rm d}}

\def\msun{M_{\odot}}

\def\ltsima{$\; \buildrel < \over \sim \;$}
\def\lsim{\lower.5ex\hbox{\ltsima}}
\def\gtsima{$\; \buildrel > \over \sim \;$}
\def\gsim{\lower.5ex\hbox{\gtsima}}

\def\Mv{M_{\rm v}}
\def\mv{m_{\rm v}}
\def\Rv{R_{\rm v}}
\def\Dv{\Delta_{\rm v}}
\def\ai{\alpha_0}
\def\ao{3}
\def\aas{\alpha_{\rm a}}
\def\rt{r_{\rm t}}

\def\rs{r_{\rm s}}
\def\rc{r_{\rm s}}
\def\ellc{\ell_{\rm s}}
\def\rhoc{\rho_{\rm s}}
\def\sigc{\sigma_{\rm s}}
\def\mf{m_{\rm fin}}
\def\mnu{\nu}
\def\Ft{F_{\rm t}}
\def\Fti{F_{{\rm t}i}}
\def\bFt{{\bf F}_{\rm t}}
\def\br{{\bf r}}
\def\bv{{\bf v}}
\def\bx{{\bf x}}

\def\la{\langle}
\def\ra{\rangle}

\def\kti{k_{{\rm t}i}}
\def\ellt{\ell_{\rm t}}
\def\elltm{\ell_{\rm tm}}
\def\rem{r_{\rm m}}
\def\alpham{\alpha_{\rm m}}
\def\rapo{r_{\rm a}}
\def\rperi{r_{\rm p}}

\def\pmb#1{\setbox0=\hbox{#1}
\kern-.025em\copy0\kern-\wd0
\kern.05em\copy0\kern-\wd0
\kern-.025em\raise.0433em\box0}
 
\def\bell{\pmb{$\ell$}}
\def\btheta{\pmb{$\theta$}}

\title[Halo cusp-core: tides in mergers]
{Galactic halo cusp-core: tidal compression in mergers}

\author[A. Dekel, J. Devor \& G. Hetzroni]
{Avishai Dekel, Jonathan Devor \& Guy Hetzroni\\
Racah Institute of Physics, The Hebrew University, Jerusalem 91904, Israel}

\begin{document}

\pagerange{\pageref{firstpage}--\pageref{lastpage}} \pubyear{2002}
\maketitle
\label{firstpage}

\begin{abstract}
We explain in simple terms how the buildup of dark haloes by merging compact 
satellites, as in the CDM cosmology, inevitably leads to an inner cusp
of density profile $\rho \propto r^{-\alpha}$ with $\alpha \gsim 1$, as seen in 
cosmological N-body simulations.  A flatter halo core with $\alpha <1$ exerts 
on the satellites tidal compression in all directions, which prevents deposit 
of stripped satellite material in the core region. This makes the satellite 
orbits decay from the radius where $\alpha \sim 1$ to the halo centre with no 
local tidal mass transfer and thus causes a rapid steepening of the inner 
profile to $\alpha >1$. These tidal effects, the resultant steepening of the 
profile to a cusp, and the stability of this cusp to tandem mergers with 
compact satellites, are demonstrated using N-body simulations. The transition 
at $\alpha \sim 1$ is then addressed using toy models in the limiting cases of 
impulse and adiabatic approximations and using tidal radii for satellites on 
radial and circular orbits. In an associated paper we address the subsequent 
slow convergence from either side to an asymptotic stable cusp with $\alpha 
\gsim 1$.  Our analysis thus implies that an inner cusp is enforced when small 
haloes are typically more compact than larger haloes, as in the CDM scenario, 
such that enough satellite material makes it intact into the inner halo and is 
deposited there.  We conclude that a necessary condition for maintaining a flat
core, as indicated by observations, is that the inner regions of the CDM
satellite haloes be puffed up by about 50\% such that when they merge into a 
bigger halo they would be disrupted outside the halo core. This puffing up 
could be due to baryonic feedback processes in small haloes, which may be 
stimulated by the tidal compression in the halo cores.
\end{abstract}

\begin{keywords}
{cosmology: theory ---
dark matter --- 
galaxies: formation ---
galaxies: haloes --- 
galaxies: interactions ---
galaxies: structure}
\end{keywords}

\section{Introduction}
\label{sec:intro}

Cosmological N-body simulations of dissipationless hierarchical clustering 
from Gaussian initial fluctuations reveal a relatively robust universal 
shape for the density profile of dark-matter haloes,
\be
\rho (r) = {\rhoc \,
           \left({r\over\rc}\right)^{-\ai} \,
	   \left(1+{r\over\rc}\right)^{\ai-\ao}
	   } \, ,
\label{eq:nfw}
\ee
where $\rc$ is a characteristic inner radius and $\rhoc$ a corresponding 
inner density.\footnote{a useful generalized functional form has been
proposed by Zhao (1996)} It has an inner ``cusp" $\propto r^{-\ai}$,
a turn-over near $\rc$, and an outer envelope of $r^{-3}$ (or perhaps somewhat
steeper) extending out to the virial radius $\Rv$.\footnote{The virial radius 
is defined by a fixed mean overdensity $\Dv$ above the universal mean, 
with $\Dv =180$ to 340, depending on time and the cosmological model.} 
Navarro, Frenk \& White (1995; 1996; 1997, hereafter NFW) found 
\equ{nfw} with $\ai \simeq 1$ 
to be a good fit to haloes in simulations
over the radius range $(0.01-1)\Rv$,
for a wide range of halo masses and for a range of hierarchical cosmological 
scenarios with different power spectra of initial fluctuations. 
Cole \& Lacey (1996) came to a similar conclusion for self-similar scenarios
with power-law power spectra, $P_k\propto k^n$ with $n=0,-1,-2$, 
in an Einstein-deSitter cosmology.
High-resolution simulations of a few individual haloes in a cosmological 
environment (Moore \etal 1998; Ghigna \etal 2000; Klypin \etal 2001)
found that the typical asymptotic cusp profile at $r\ll \rc$ is
sometimes somewhat steeper, closer to $\ai \simeq 1.5$. 
A careful convergence analysis by Power \etal (2002), who explored the
robustness to numerical errors, 
found for the standard $\Lambda$CDM cosmology that $\ai$ reaches a slope
shallower than 1.2 at their innermost resolved point of $r \sim 0.005\Rv$. 
Given these uncertainties regarding the exact slope and its extent, 
the simulations have established the formation of a cusp with a 
characteristic slope in the range $1 \leq \ai \leq 1.5$.
However, a basic theoretical understanding of its origin is still lacking.

An even more intriguing puzzle is introduced by observations of low
surface-brightness (LSB) or dwarf galaxies,
whose centres are dominated by their dark matter haloes, which indicate
that at least in some cases their actual inner halo density profiles are 
close to flat cores, with $\ai \simeq 0$ (e.g., van den Bosch \etal 2000;
de Block \etal 2001; Marchesini \etal 2002).
Similar cores may be present in other galaxies (Salucci \& Burkert 2000;
Salucci 2001; Borriello \& Salucci 2001).
This seems to introduce a severe challenge to the CDM cosmological paradigm.
Attempts to turn a cusp into a core by direct stellar feedback effects 
in the present large haloes, which looked promising at a first sight
(Navarro, Eke \& Frenk 1996), seem not to work in a straightforward way
(e.g., Geyer \& Burkert 2001; Gnedin \& Zhao 2002).

In order to make progress in resolving the core problem, we find it useful
to first try to understand in simple basic terms the origin of the universal 
cusp in the gravitational N-body simulations of cold dark matter. 
This should provide us with a tool for addressing the formation and survival
of flat cores by some non-gravitational mechanism, perhaps by baryonic feedback 
processes still within the hierarchical CDM framework.

We now return to the issue of dark-halo profiles in dissipationless 
simulations. 
The outer slope of $r^{-3}$ (and steeper) may possibly be explained in terms 
of violent relaxation (e.g., Barnes \& Hernquist 1991; 
Pearce, Thomas \& Couchman 1993 and references therein).
In general, any
finite system would tend to have a steep density fall off at large radii
due to diffusion of particles outwards.
Secondary spherical infall is expected to produce a profile closer to
the profile of an isothermal sphere, $\rho \propto r^{-2}$, which may
explain the behavior in the intermediate regions of the halo, but is
too steep to explain the flatter inner cusp of $\alpha \leq 1.5$
(Lokas \& Hoffman 2000 and references therein).
Thus, none of the above mechanisms provides a natural explanation for 
the characteristic cusp of $\ai \gsim 1$.

By following the evolution in
cosmological N-body simulations of hierarchical clustering scenarios, 
we know that haloes are largely built
up by a sequence of mergers of smaller haloes
(e.g., Klypin \etal 1999b; Moore \etal 1999a; Springel \etal 2001).\footnote{
Although claims were made that a substantial
fraction of the mass may actually be added by smooth accretion rather than by
mergers, a careful inspection of high resolution CDM simulations shows
that this is a wrong impression arising from difficulties in identifying
the merging clumps in many cases (e.g., Wechsler, Dekel \etal, in
preparation).}
In a typical merger, a bound 
satellite halo spirals into the centre of the larger halo due to gravity 
and dynamical friction. The satellite 
gradually transfers mass into the host halo due to tidal stripping
or by eventually melting into the halo inner region. 
This process is likely to have an important effect in shaping up the 
density profile.
Indeed, Syer \& White (1998), Nusser \& Sheth (1999) and Subramanian,
Cen \& Ostriker (2000) argued, using certain
simple models and simulations of a sequence of mergers, that the buildup by 
mergers may naturally lead to a stable profile. 
However, they find their predicted profile to be quite sensitive to the power 
spectrum of fluctuations and to allow an inner slope of $\ai <1$, 
in conflict with the robust result of the cosmological simulations. 
In fact, when trying to repeat the Syer \& White analysis using 
their simplified modeling of the stripping process but with higher resolution, 
we find that in the long run the profile does not really converge to a 
stable cusp but rather continues to steepen slowly towards $\ai=3$.
Either way, it seems that something is not adequate in the simplified
model adopted to describe the mass transfer from the satellite to the halo
in these studies.

We re-visit here the buildup of halo profile by merging 
satellites and gain an encouraging new insight.
We add two important new ingredients to the tidal effects.
In the current paper, we argue that for a flat mean halo density profile 
with a logarithmic slope
$\alpha \leq 1$ the tidal effects on typical satellites induce 
three-dimensional compression with no local mass deposit, 
which results in a rapid steepening of the inner
profile to $\alpha > 1$.
In an associated paper (Dekel \etal 2003)
we derive a useful prescription for tidal mass transfer at $\alpha > 1$,
and obtain higher deposit efficiency at higher $\alpha$.  
We then show that this tends to flatten steep profiles with large
$\alpha$ and thus slowly leads to an asymptotic fixed point at a certain
$\alpha = \aas \gsim 1$.

The key idea in the current paper is that the local tidal mass transfer 
from the merging satellite to the host halo practically stops when the 
satellite's orbit has decayed into
a core region where the local logarithmic slope is flat, $\alpha(r) \leq 1$. 
As a result, the satellite 
continues to sink in due to dynamical friction without mass loss
until it settles in the halo centre.
This inevitably causes a general steepening of the core profile
towards $\alpha > 1$, thus explaining the asymptotic inner slope of
the cusp in the NFW profile as seen in the cosmological simulations.
We demonstrate this behavior using merger N-body simulations
and interpret it using crude analytic approximations in idealized cases.

In \se{tide} we address the $\alpha$ dependence of the tidal force and
highlight the three-dimensional tidal compression in a core of $\alpha < 1$.
In \se{sim1} we describe the N-body merger simulations.
In \se{sim2} we analyze the evolution of halo
density profile in these simulations, where the small satellite
is more compact than the more massive halo, as in a typical
hierarchical clustering scenario (e.g. $\Lambda$CDM).
In \se{limits} we qualitatively evaluate the transition in mass-transfer 
efficiency at $\alpha \sim 1$ in the extreme limits of impulse and adiabatic 
tidal effects along typical satellite orbits.
In \se{tidal_radius} 
we continue to study the $\alpha$ dependence of the mass transfer
using the tidal radii for mergers on radial and circular orbits.
In \se{asymp} we summarize the analysis leading from either side
to an asymptotic cusp $\alpha \rightarrow \aas \gsim 1$.
In \se{core} we demonstrate how the change from a core to a cusp
could be avoided when the merging satellites are slightly puffed up,
and discuss a possible scenario for puffing up small haloes
by baryonic feedback, perhaps assisted by tidal compression.
In \se{conc} we discuss our results and put them in a broader perspective.

\section{Tidal Force: Compression and Stretching}
\label{sec:tide}

\begin{figure*}
\vskip 4.5 cm
{\includegraphics{f1.ps}}
\caption{
An illustration of the tidal acceleration at a given satellite shell 
of radius $\ell\ll r$, for a given $\bar\rho(r)$
and for different values of halo density-profile slope $\alpha$ 
[\equ{tide_harmonic} or \equ{tide}].
Along the line connecting the centres of mass ($\hat{\bell}_1=\hat{\br}$)
the tides stretch outwards for $\alpha>1$ and compress inwards for $\alpha<1$.
Along the perpendicular directions, $\bell_2$ or $\bell_3$, there is
always compression, independent of $\alpha$.
}
\label{fig:tide}
\end{figure*}

A straightforward analysis indicates that
the tidal forces exerted by a dark halo on a satellite orbiting in it
depend not only on the distance from the halo centre and the halo mass
within that radius, but also on the local slope of the halo density profile.
This dependence, which is often overlooked, affects the way mass is 
transfered from the satellite to the halo, and may provide a clue for
the origin of the characteristic cusp of $\alpha \geq 1$ in haloes.

We consider a spherical halo of mass profile $M(r)$ and virial mass
$\Mv$.  The {\it mean} density in a sphere of radius $r$ is
$\bar\rho(r) \propto M(r)/r^3$.
A useful quantity in describing 
the tidal forces exerted by this halo is its local logarithmic slope,
\be
\alpha(r) \equiv - {d\ln \bar\rho \over d\ln r} \,,
\ee
such that locally $\bar\rho \propto r^{-\alpha}$.
We assume that $\alpha$ is either constant or monotonically increasing as
a function of $r$, with values in the range $0 \leq \alpha \leq 3$.
The extreme values of $\alpha=0$ and 3 correspond to a constant-density
halo and a point mass respectively.
Note that if the profile inside $r$ is a power law, then the local and
mean density profiles have the same logarithmic slope. In general,
they are related via $\rho(r) =[1-\alpha(r)/3] \bar{\rho}(r)$,
but they do not necessarily have the same slope at a given $r$.
The slope of $\rho(r)$ is equal to or larger than the slope of $\bar\rho(r)$
at any $r$.
The following analysis refers to $\alpha$ as the slope of $\bar\rho(r)$.

In the analytic part of our analysis
we assume that the density profile of the original halo mass is fixed in time
throughout the duration of the merger while the mass torn from the satellite 
is gradually being added to the halo.
This is confirmed to be a reasonable approximation in our N-body
merger simulations, where the mass ratio is 1:10 (e.g. \Fig{alpha_r}).

We then consider a satellite of mass $\mv \ll \Mv$, moving under the
gravity and dynamical friction exerted by the halo, 
when the satellite's centre of mass is at position
$\br$ as measured from the halo centre.
The tidal acceleration exerted by the halo mass distribution on a satellite
particle at position vector $\bell$ relative to the satellite centre of mass
is obtained by transforming the gravitational attraction exerted by the halo
on the particle into the accelerated (non-rotating)\footnote{in some cases it 
is useful to perform the analysis in a rotating frame, see \se{tidal_radius}}
rest frame of the satellite,
namely, by subtracting the acceleration of the satellite centre of mass
relative to the halo,
\be
{\bFt} = -{G M(|\br+\bell|)\, (\br+\bell) \over |\br+\bell|^3}
         +{G M(r)\, \br \over r^3} \,.
\ee
In the tidal limit $\ell \ll r$, this yields to first order in $\ell/r$
\be
{\bFt} = {G\ell \over r^3}\,
         \left( [3M(r) -M'(r)\,r]\, (\hat{\br} \cdot \hat{\bell})\, \hat{\br}
         \,-\, M(r)\, \hat{\bell} \right) \,,
\ee
where $M'(r) \equiv dM/dr$.
This expression can be simplified using the definition of $\alpha(r)$, which
gives $M'(r)\,r = [3-\alpha(r)] M(r)$.
Using Cartesian coordinates about the satellite centre, where
$\bell = (\ell_1,\ell_2,\ell_3)$
and $\bell_1$ lies along $\br$, we get
\be
{\bFt} = {GM(r) \over r^3}\,
\left( [\alpha(r)-1]\,\bell_1\, -\bell_2 -\bell_3 \right) \,.
\label{eq:tide_harmonic}
\ee
Alternatively, moving to polar coordinates, where $\theta$
is the angle between $\bell$ and $\br$ and
$\hat{\br}=\cos\theta\, \hat{\bell} -\sin\theta\, \hat{\btheta}$,
we obtain
\be
{\bFt} =
{GM(r) \ell \over r^3}\,
\left(\, [\alpha(r)\, \cos^2\theta -1]\ \hat{\bell}
\ - \
\alpha(r) \sin\theta \cos\theta \ \hat{\btheta} \, \right)
\,.
\label{eq:tide}
\ee

\Fig{tide} illustrates the tidal acceleration at a given satellite shell
of radius $\ell$ in a plane that includes the position vector $\br$ connecting
the centres of mass,
for given $r$ and $M(r)$ and for different slopes $\alpha=3$, 2, 1 and 0.
The tidal components along $\bell_2$ and $\bell_3$,
perpendicular to $\br$, are always negative, namely, they exert compression,
and they do not explicitly depend on $\alpha$.
Stretching outwards may occur only along $\pm \br$.
Unlike the common notion of tidal forces as pulling outwards, the average
force on a sphere of radius $\ell$ is actually inwards:
$\la \bFt \ra =(GM/r^3) (\alpha/3-1)\bell$.
It vanishes for a point-mass perturber, $\alpha=3$, and it obtains
a maximum amplitude in a core where $\alpha=0$.

According to \equ{tide}, 
the maximum radial tidal force outwards (towards $+\hat{\bell}$)
is always obtained along the line connecting the centres of mass, $\pm \br$.
According to \equ{tide_harmonic}, and as illustrated in \Fig{tide},
the pull outwards is maximal 
in the limit where the tides are exerted by a point-mass halo, $\alpha=3$,
valid in the outer regions of typical haloes.
For flatter halo profile slopes, the tidal stretching
becomes weaker in proportion to $(\alpha-1)$, until it vanishes at $\alpha=1$
and reverses direction into compression for $\alpha<1$.
This direction reversal of the tides is often overlooked, because the common
context of tides is a perturbation by a concentrated mass distribution.
It is a simple result of the fact that the amplitude of the gravitational
attraction by the halo, $GM(r)/r^2$, is not a decreasing function of $r$
where the halo mass profile is rising rapidly enough (faster than $M \propto
r^2$).
Thus, while for $\alpha>1$ there is always a tidal component pulling outwards,
for $\alpha < 1$ the tidal forces are of compression for any $\theta$,
namely everywhere in the satellite.  In the limit of a constant-density core,
$\alpha=0$, the tides induce symmetric compression in all three dimensions.

We note that the critical slope of $\alpha=1$ is the asymptotic inner slope of
the cusp in the NFW profile.  We suggest that this is not a coincidence.
The key idea is that if the local tidal mass transfer from the satellite
to the halo stops when the satellite's orbit has decayed into
a core region where $\alpha(r) \leq 1$, the satellite would continue to sink 
in due to dynamical friction until it settles in the halo centre.
This would inevitably cause a general steepening of the core profile towards
$\alpha > 1$.
For this scenario to be valid one should verify that, indeed,
there is only little local tidal transfer of mass from the satellite to 
the halo in a region where $\alpha \leq 1$.
We first demonstrate this using N-body simulations,
and then apply crude analytic approximations in simple
cases in order to gain better qualitative understanding of this effect.

\section{Merger Simulations}
\label{sec:sim1}

For the purpose of investigating the proposed scenario,
we run and analyze N-body simulations of isolated mergers between a
large halo and a satellite halo of mass ratio $m/M=0.1$, and then
try to interpret them using analytic toy models.
We use the Tree code by Mihos \& Hernquist (1996 and references
therein) with dark-matter haloes only (no gaseous disks).
In our default suit of simulations, the large host halo is represented
by $N=10^5$ equal-mass particles and the satellite by $n=10^4$ particles.
We test the sensitivity of our results to resolution by repeating one
case with $N+n=0.55\times 10^6$ particles and a correspondingly
higher force resolution.
The simulation units were chosen quite arbitrarily to be: length 
$3.5$kpc, mass $5.6\times 10^{10}\msun$, and time $13.06$Myr.
The force softening length in the default case is 0.08 units, i.e. $0.28$kpc.
The units can be properly scaled up or down according to desire.
In most cases we refer hereafter to distances and masses in the simulation  
units, but to time in Myr. 

\begin{figure}
\vskip 7cm
{\includegraphics{f2.ps}}
\caption{
The halo profile in the simulation represented by the
logarithmic slope $\alpha$ of the mean density profile
as a function of radius $r$ (in units of 3.5kpc).
Shown are the halo profile at the initial time (solid) and the profiles
of the original halo mass at the final times of the radial (long dash)
and circular (short dash) merger simulations.
}
\label{fig:alpha_r}
\end{figure}

The initial halo density profile, as measured in the unperturbed initial
conditions, is a truncated isothermal sphere with a flat core,
\be 
\rho(r) = {\rhoc\, e^{-(r/\rt)^2} \over 1 + (r/\rc)^2} \,,
\label{eq:halo_prof}
\ee 
with $\rhoc=10.36$, $\rc=1$ and $\rt=10$. 
The density at the characteristic radius $\rc$ is thus $\rho(\rc)=5.13$. 
The internal velocities are constructed to fulfill the isotropic Jeans
equation which ensures an equilibrium
configuration as discussed in Mihos \& Hernquist (1996).
When run in isolation, the halo profile has been tested to be very
stable for many dynamical times.
As shown in \Fig{alpha_r}, the logarithmic slope of the density profile
spans the range of interest between $\alpha=0$ and 3, and
its variation as a function of radius can be
described to a good approximation by $\alpha(r) \approx 1.73 \log r +0.67$
throughout the range $0.3 \leq \alpha \leq 2.9$.
The initial halo density profile will be shown in \Fig{prof_compact} below  
in comparison with the post-merger profile. 
It resembles the generalized NFW profile 
of \equ{nfw} with a core of $\alpha \ll 1$.

\begin{figure*}
\vskip 16.5 cm
\includegraphics{f3a.ps}
\includegraphics{f3b.ps}
\includegraphics{f3c.ps}
\caption{
The $10^4$ satellite particles in 6 snapshots during the radial (left),
circular (right) and elongated (bottom) mergers, 
projected onto the orbital plane.
The $10^5$ live halo particles are not shown.
The centre of mass is at the origin and the dot
marks the temporary halo maximum density.
The circle is of radius $r=20\simeq 2\rt$ about the halo maximum
density,
corresponding to where the initial halo practically ends and where
the satellite is at the onset of the simulation.
The three thirds of the mass, in concentric shells about
the satellite bound centre at each time, are marked by different colors.
}
\label{fig:snaps}
\end{figure*}

\begin{figure*}
\vskip 9.5cm
{\includegraphics{f4a.ps}}
{\includegraphics{f4b.ps}}
{\includegraphics{f4c.ps}}
{\includegraphics{f4d.ps}}
\caption{
Time evolution of the satellite spherical mass profile during the radial
merger (left) and the circular merger (right).
Shown in the bottom panels are the mean radii of concentric spherical
shells
about the momentary satellite centre, each encompassing a given
fraction of the original satellite mass, as labeled on the right.
The top panels show the time evolution of position $r$ of the satellite
centre relative to the halo centre (dotted), and the
corresponding local slope $\alpha(r)$ (solid).
For the radial merger, the times of halo-centre crossing are marked
in the bottom panel by vertical lines.
For the circular merger, representative values of $\alpha(r)$ along the
orbit are marked in the bottom panel.
The stripping of a shell can be crudely identified by a rapid increase
of its radius.
Lack of stripping and slight overall contraction of bound shells
is noticed whenever the satellite enters the halo core.
}
\label{fig:radii}
\end{figure*}

The satellite initial density profile is fit by a Hernquist profile,
\be
\sigma(\ell) = {\sigc \over (\ell/\ellc)\, [1+(\ell/\ellc)]^3} \,,
\label{eq:sat_prof}
\ee
with the default choice $\sigc=19.2$ and $\ellc=1$ defining our
typical ``compact" satellite.
The initial satellite density profile can also be seen in \Fig{prof_compact}
below. 
In the inner region it convergence to the NFW profile, $\alpha=1$.
If we fit this Hernquist profile with an NFW profile by matching
the characteristic radii where the local logarithmic slope is $-2$,
we find that the radius corresponding to the characteristic
radius of the NFW profile is $\ell=\ellc/2$. At this radius the
density is $16\sigc/27$, which for our default satellite is 
$\sigma(\ellc/2)=11.38$.

The satellite parameters were chosen to roughly mimic a typical 
compact satellite according to the distribution of halo properties
in the $\Lambda$CDM scenario.  In a hierarchical clustering scenario with a 
fluctuation power spectrum $P_k \prop k^n$, the 
halo characteristic radii and densities scale like
$\ellc/\rc \propto m^{(1+\mnu)/3}$ and $\sigc/\rhoc \propto  m^{-\mnu}$,
with $\nu = (3+n)/2$. 
For $\Lambda$CDM on galactic scales one has $\mnu \simeq 0.33$.
This is confirmed by cosmological simulations (NFW; Bullock \etal 2001a),
where the typical halo profile is NFW.
The haloes of lower masses are thus typically more compact; they
tend to have lower characteristic
radii and higher corresponding mean densities within these radii
(corresponding to higher virial concentration parameters).
For CDM haloes of mass ratio $m/M=1/10$,
the typical ratio of characteristic radii and corresponding densities
are expected to be roughly $0.4$ and $2.1$.  
The corresponding ratios in the initial conditions of our simulations
are approximately $0.5$ and $2.2$, 
providing a reasonable match. 
It may also be interesting to note that the satellite mean density interior 
to $\ellc/2$ (or $\ellc$) is about $3.9$ (or $1.09$) times the halo mean 
density interior to $\rc$.

We simulated three cases of initial merger orbits:
a radial orbit, a circular orbit, and an elongated orbit with
an initial pericentre to apocentre ratio of $\rperi/\rapo \simeq 1/6$.
The unperturbed satellite is put initially at $r=20$, i.e.
at about $2\rt$, where the initial circular period is about $230$Myr
(or about $17.6$ simulation time units). 
In the circular and radial cases the magnitude of the initial satellite
velocity was set to
equal the circular velocity of the halo at that radius, namely a bound
orbit with an orbital kinetic energy that equals half the absolute value of
the total energy.
For the elongated orbit the initial tangential velocity was about one half of
the circular velocity at $r=20$, in fact $v/v_{\rm c}=0.49$.
Each merger has been followed until the satellite's bound core has
practically settled at the halo centre.
\Fig{snaps} shows the satellite mass distribution in 6 snapshots
during the merger process in the three cases, projected onto (one of) the
orbital planes. 

The radial-orbit merger has been followed for more than 200Myr 
(about 15 time units), with equal steps of 1.306Myr (0.1 time units)  
between output times.
The satellite is stretched and stripped along the line connecting the
centres of mass
and the stripping produces the familiar bridge-and-tail tidal structure.
An overall shrinking is seen near the first centre crossing, followed
by a three-dimensional re-bounce and significant mass loss about the
following apocentre (discussed in \se{limits} below).
The stripped material shows a sequence of arc-like caustic structures
corresponding to each apocentre, reflecting the initial confinement in
phase space.
The oscillations of the
satellite remnant about the halo centre decay rapidly due to dynamical
friction until it becomes confined to the halo core after about $115$Myr,
and it practically melts into the halo central region after $\sim
130-140$Myr.
The final distribution of satellite mass extends quite smoothly
about the halo centre in a puffy ellipsoid; it is moderately prolate,
with the major axis parallel to the original merger line.

The circular-orbit merger has been followed for 836Myr (64 time units) 
with 81 output times spaced by 10.45Myr.
The satellite gradually spirals inwards due to dynamical friction.
The continuous tidal stripping process in the outer parts is obvious.
Particles escape from the satellite in the two directions roughly along
the line connecting the centres of mass of halo and satellite, and
produce the familiar trailing and leading tidal tails because of the
differential rotation in the halo potential.
The remaining bound satellite seems to be slightly elongated in a
direction not far from the line connecting the centres of mass,
crudely reflecting the shape of the Roche lobe (see \se{tidal_radius} below).
After entering the halo core, at about $380$Myr,
the dynamical friction weakens and
the satellite orbit continues to decay very slowly towards the halo
centre.
The final distribution of satellite mass extends quite smoothly
in an oblate configuration about the halo centre.
               
In the elongated-orbit merger,
the satellite oscillates about the halo centre through a sequence of pericentre
and apocentre passages.
The shape of the orbit remains roughly constant while it shrinks in scale;
when measuring the ratio of pericentre to the following apocentre it is
$\rperi/\rapo \sim 1/3.5$, and when measuring the ratio of pericentre to the
preceding apocentre it is $\rperi/\rapo \sim 1/3.5$, but these ratios
remain roughly the same for all detectable pericentres.
The oscillations decay due to dynamical
friction until the satellite becomes confined to the halo core
after $\sim 125$Myr and 5 pericentre passages.
By the first pericentre, the satellite is already stretched and stripped
along its orbit, while it is temporarily shrunk in the perpendicular 
direction.
This is followed by a re-bounce and significant mass loss about the
following apocentre.  
The visual impression confirms the notion that the particles that are
torn away near a pericentre radius continue on orbits that reflect on average
the satellite orbit at the time of stripping, while the bound remnant is
gradually sinking into smaller radii due to continuing dynamical friction
(see \se{orbits} below).
For example, in the snapshots corresponding to the second and third apocentres
we clearly see a large amount of stripped satellite material spread about the
location of the previous apocentre.
We can crudely say, in all the mergers simulated,
that mass is stripped near pericentre and is practically
``deposited" about the following apocentre radius.
This crudely validates a toy-model concept that each
satellite shell $\ell$ is being practically deposited at a certain halo
radius $r$.
The final distribution of satellite mass extends quite smoothly
about the halo centre in a puffy oblate ellipsoid that looks quite symmetric
in the orbital plane.

\begin{figure*}
\vskip 13.4cm
{\includegraphics{f5a.ps}}
{\includegraphics{f5b.ps}}
{\includegraphics{f5c.ps}}
{\includegraphics{f5d.ps}}
\caption{
Halo density profile before (solid) and after (dashed) the simulated mergers
with a compact satellite.
The initial halo core becomes a cusp for all types of merger orbits (specified
within each panel).
A significant fraction of the satellite settles intact
at the halo centre without depositing mass near $\alpha \lsim 1$,
causing the steepening of the core into a cusp.
The error bars refer to Poisson errors in the $r$ bins, and they are
roughly spaced by the bin size.
The thin dotted curves mark for reference
the initial satellite profile (lower) and a
straightforward sum of the initial profiles of halo and satellite (upper).
The bottom-right panel corresponds to the same elongated-orbit merger
as in the bottom-left panel, but simulated with 5 times more particles
($0.55\times10^6$)
and a softening length smaller by a factor of $5^{1/3}$ accordingly.
It demonstrates that our default resolution is adequate.
}
\label{fig:prof_compact}
\end{figure*}

\Fig{alpha_r} displayed earlier 
also shows the $\alpha(r)$ profiles
of the original halo material at the final times after the mergers.
The relatively small changes in this profile compared to the changes
in the profile of the total halo mass (original halo plus stripped
satellite material) indicate that the assumption
of a fixed halo to which the satellite mass is being added can serve
for a crude toy model.

\section{Profile Buildup in the Simulations}
\label{sec:sim2}

\Fig{radii} describes the time evolution of the satellite spherical
mass profile during
the mergers by showing the mean radii of concentric spherical shells
about the satellite maximum-density centre, each encompassing a given
fraction of the satellite mass (and not necessarily the same
population of particles at different times).

In the circular merger case, to a good
approximation, the satellite position $r$ is a monotonically decreasing
function of time, and so is $\alpha(r)$, as shown at the top and marked
along the time axis of \Fig{radii}.
The stripping moment of each shell is marked by the onset of a steep rise
in the corresponding curve.  Proceeding from the outside shells inwards,
the stripping point of each shell can be identified with
a specific halo radius $r$ and a corresponding $\alpha$, moving down
monotonically from $\alpha \sim 3$ to flatter slopes.
However, as the stripping point passes beyond the radius where
$\alpha \sim 1$  (after $\sim 380$Myr), 
when the satellite is left with less than $\sim 30\%$ of its original mass,
the stripping slows down and eventually stops. The 20\% and inner
shells are never really stripped, 
while the satellite's orbit continues to
decay very slowly towards the halo centre.

In the radial merger case the satellite oscillates about $r=0$.
It crosses the halo core back and forth a few times while the oscillation
amplitude is decaying due to dynamical friction until the bound remnant
settles at the halo centre. Center crossings occur near the times 
22, 62, 83, 99, 110Myr and so on.
The satellite enters the core region in every oscillation
a couple of Myr before it crosses the halo centre. 
Overall contraction of bound satellite shells seems to start roughly at these
times, as indicated by the reversal of the tidal forces near $\alpha =1$ 
(\se{tide}).
Each major contraction is followed by a re-bounce as the
satellite exits the core towards apocentre in the opposite side,
which results in overall expansion and stripping of the outer shells.
                    
After about 90Myr the satellite orbit becomes confined to inside the
region where $\alpha < 1.5$, and after about 107Myr it is confined to
the inner halo where $\alpha < 1$.
Once the satellite becomes confined to this halo core,
there is no apparent overall shell expansion anymore, indicating
that stripping has stopped.

In the elongated merger case (not shown to avoid redundancy),
pericentre passages are identified five times, and the behavior is
qualitatively similar to the radial merger case.
Overall contraction of bound satellite shells seems to start roughly at
these times.
Each major contraction is followed by a re-bounce as the 
satellite moves towards apocentre, 
which results in overall expansion and stripping of the outer shells.
Once the satellite becomes confined to the inner halo where $\alpha < 1$,
after $\sim125$Myr and 5 apocentres,
there is no apparent overall shell expansion anymore, indicating
that stripping has stopped.

It is important to
note that the halo radius $r$ where $\alpha=1$ is more than three times
larger than the radius $\ell$ of the satellite's 20\% shell,
indicating that the cease of stripping occurs significantly before the
centre of the bound satellite remnant coincides with the halo centre.

\Fig{prof_compact} addresses the cusp formation straightforwardly
by showing the density profile of the halo before and after the
merger in the three different cases. 
For the standard, compact satellite, either on radial, circular or elongated 
orbit, the figure demonstrates the inevitable steepening of the profile
in the core region, starting near the core boundary where $\alpha \simeq 1$.
In the case of radial merger, the slight depletion of final density
in the region near $\alpha \sim 1$, compared to the slight increment in
halo density at larger radii, is consistent with no mass transfer in this
region while the orbit of the remaining satellite continues to decay into
smaller radii. A similar effect is also seen in the circular case, though it is
somewhat weaker.
Not surprisingly, the steepening seen at $\alpha \lsim 1$ in the elongated
merger case is not very different from the steepening seen in the
two other cases.

In order to evaluate the sensitivity of our results to the resolution of
the simulation, we ran a case identical to the merger on an elongated orbit
described above, but now with 5 times more particles ($N+n=0.55\times 10^6$)
and a softening length smaller by a factor of $5^{1/3}$ accordingly. 
The results are found to be practically identical in all respects,
including the decay rate of the satellite radius within the halo,
the mass loss from the satellite, and the final halo density profile.
The latter is shown in the bottom-right panel of \Fig{prof_compact}.
This indicates that the resolution of our simulations is adequate 
in the range of radii of relevance, well inside the core/cusp region
and down to below $0.2\rs$.

\begin{figure}
\vskip 7.0cm
{\includegraphics{f6.ps}}
\caption{
Halo density profile after tandem mergers
with compact satellites of mass ratio 1:10 on random elongated orbits.
The initial halo profile, with a core, is shown for comparison (solid line).
After each merger, the total mass of the halo is scaled down back to
the original mass. We see that the cusp which forms in the first merger
remains stable under the subsequent mergers.
}
\label{fig:prof_tandem}
\end{figure}

We thus learn that one single merger with a 1:10 compact satellite, 
typical of the
$\Lambda$CDM cosmology, is enough for turning a core into a cusp of $\alpha
\gsim 1$. How stable is this cusp under further mergers with similar
satellites? To test this, we performed a series of mergers following each
other. The main progenitor halo in each merger
is taken to be the outcome of the previous
merger, except that its total mass is scaled down back to the original halo 
mass (such that all the density profiles can be directly compared).
We achieve this by letting each simulation involve the same numbers of
$N$ halo particles and $n$ satellite particles (each of a fixed mass $m$),
where the halo particles are selected at random from the $N+n$ particles
of the halo produced in the previous merger.
In these tandem mergers we used $N=20,000$ and $n=2,000$.
The halo profile in the range of interest is found to remain stable 
for at least several hundred Myr when the halo is run in isolation.
The satellite starts with the same profile as the original compact satellite. 
It is put at the same
distance of $r=20$ as before but in an elongated orbit of a random spatial 
orientation. The initial satellite velocity is tangential to the line
connecting the centres of mass of halo and satellite, with an amplitude 
in units of the circular velocity at $r=20$ 
chosen at random in the range $v/v_{\rm c}=0.2-0.6$ 
(compared to the typical case simulated before, where $v/v_{\rm c}=0.49$ 
led to peri/apocentre ratio of roughly 1:6).  
Each merger was followed for 325Myr before the following merger started.
The density profiles after each of the first 5 mergers are shown
in \Fig{prof_tandem}. We see that
the first merger reproduces a cusp very similar to the cusp produced
when the merger was simulated with $N=100,000$ (and with $N=500,000$),
\Fig{prof_compact},
indicating that even $N=20,000$ is adequate for crude results in the regime
of interest here.
We see that inside the initial core radius the
cusp is stable at a slope of $\alpha \simeq 1.5$, 
with no systematic tendency to deviate
from the power-law profile resulting already after the first merger.
At the radius where $\alpha \gsim 1.5$ the profile flattens, and 
the amplitude becomes
lower partly because of the renormalization of the total mass between
every two mergers.

We conclude that 
the NFW inner slope of $\alpha =1$ is indeed a robust lower bound,
as indicated by the reversal of the tidal forces there (\se{tide}).
Any flatter density core does not survive as long as bound satellites bring
enough mass into the inner halo and deposit it near the halo centre.
This provides a simple explanation for why the cusp slopes seen
in cosmological N-body simulations of CDM models are typically $\alpha > 1$.
It implies that the only way which may possibly enable a flatter core 
is by suppressing the settling of satellites in the core, as demonstrated 
by simulations with puffy satellites (\se{core}).

Next, we continue our effort to understand the tidal mass transfer 
seen in the simulations. We build upon the $\alpha$ dependence of the tidal 
forces discussed in \se{tide}, and try to evaluate the resultant mass transfer
using simple dynamical considerations and idealized toy models.

\section{The Impulse and Adiabatic Limits}
\label{sec:limits}

In this section we analyze the tidal mass transfer from satellite to halo
by referring to particles in orbits within the satellite.
The $\alpha$ dependence of the tidal process in the central
regions of a halo can be qualitatively evaluated in two extreme limits.
If the tidal forces vary on a time scale shorter than the particle
orbital periods within the satellite, then one can use the impulse 
approximation (Spitzer 1958; Binney \& Tremaine 1987, hereafter BT, \S 7.2) 
to estimate the energy input into the 
satellite and the resulting mass loss.
If the tidal forces vary on a time scale longer than the orbital periods
within the satellite, then one can appeal to adiabatic invariants along the
particle orbits (e.g., BT \S 3.7)
in order to estimate the gradual energy change of the satellite particles,
the associated structural changes in the satellite, and the escape rate.

\subsection{Orbits, stripping and deposit}
\label{sec:orbits}

The validity of each approximation depends on the nature of the orbit of 
the satellite within the halo.
Along a typical orbit, the satellite distance from the halo centre oscillates
periodically between the radii of apocentre $\rapo$ and pericentre $\rperi$, 
while their amplitudes gradually decay due to dynamical friction.
Ghigna \etal (1998) studied the distribution of satellite orbits in a
high-resolution N-body simulation of a cluster emerging from a CDM cosmological
background (mostly in an extended range where $\alpha \sim 2$). 
They found that the median ratio $\rapo/\rperi$
is 6:1, with about 25\% of the orbits more eccentric than 10:1,
and concluded that radial orbits are common while circular orbits are rare.
The expected more rapid tidal disruption of satellites on radial orbits
may indicate that the actual initial distribution of orbits tended
even more towards radial orbits.

Ghigna \etal (1998) also demonstrated that the tidal radii of the satellites,
$\ellt$, are consistent with being determined near pericentre, under the 
general resonance condition $\bar\rho(\rperi) \sim \bar\sigma(\ellt)$, where
$\bar\sigma(\ell) \propto m(\ell)/\ell^3$ is the mean density profile
of the satellite. Particles that escape from the satellite can be assumed,
on average, to continue on an orbit about the halo centre with apocentre 
and pericentre radii ``frozen" at their values near the time of escape,
suffering no further decay due to dynamical friction. Since a particle spends
most of its time near the apocentre of its orbit, we can vaguely say that the 
escapers are effectively ``deposited" in the halo near the apocentre radius 
valid at the time of escape.  This effect is clearly seen in our simulation
of the radial and elongated merger orbits, e.g.,
\Fig{snaps}, where a large cloud of stripped particles is seen left
behind at each successive apocentre.

Very eccentric orbits may involve ``penetrating" encounters near pericentre,
where $\rperi < \ell$ for $\ell$ a characteristic radius of the satellite.
In this case the tidal approximation becomes invalid. The tides are weakened 
when the centre of the perturber lies inside the satellite, until
they vanish when the satellite centre coincides with the halo centre. 
It can be shown that the effect of a fast penetrating encounter
is comparable to the effect of an encounter with $\rperi \sim \ell$ as computed
using the impulse approximation in the tidal limit (BT \S 7.2.e). 
In a slow penetrating situation the tidal effects simply weaken gradually as 
$r \rightarrow 0$.

\subsection{The impulse limit}
\label{sec:imp}

The impulse approximation may be partly valid for satellites on elongated
orbits during their first quick pericentre passages.
As long as $\alpha>1$ along the orbit, the tidal force, 
$\Ft \propto \bar\rho(r)$, has a peaked maximum at the $\sim\rperi$
vicinity of the pericentre, where the satellite spends a small fraction of its
orbital period. Since $\bar\rho(\rperi) \sim \bar\sigma(\ellt)$,
and since the typical orbital period is inversely proportional to the
square-root of the density,
the satellite orbital period is comparable to the internal orbital
periods of outer satellite particles. Therefore, the impulse approximation
may be partly valid for such particles near their apocentres within 
the satellite.

In the impulse approximation,
each satellite particle is assumed to obtain an instantaneous
velocity kick $\Delta\bv$, which is the integral $\int \bFt \d t$ 
over the short duration 
of effective tides near pericentre along the satellite orbit.
The energy change per unit mass
of a particle moving with momentary velocity $\bv$
is thus $\Delta E=(1/2)(\bv+\Delta\bv)^2-(1/2)\bv^2$, 
which consists of linear and quadratic terms in $\Delta\bv$, namely
\be
\Delta E=\bv\cdot\Delta\bv + (1/2)(\Delta\bv)^2 \,. 
\ee
Some particles may be kicked by this impulse to above the escape velocity 
at their position and become unbound. 

If the satellite is spherical and the particle motions outwards and inwards
are symmetric, a kick inwards is as effective in feeding energy into the 
satellite as a kick outwards.
The contribution of the linear term to the total energy vanishes
by the symmetry of incoming and outgoing particles, but the linear term
may have an important contribution for the escapers which come from the 
high-end tail of the $\Delta E$ distribution.

The dominant tidal kick $\Delta\bv$ is due to the systematic force towards the
satellite centre along the direction perpendicular to the orbital plane,
where the tidal force is $\alpha$-independent and persistently inwards
[\equ{tide_harmonic}].
A somewhat smaller kick is obtained when integrating the force component
along the position vector of the satellite ($\br)$ at pericentre 
(it is a net kick inwards if $\alpha < 2$), while the contribution along
the position vector at apocentre tends to average out when integrated along
the orbit (BT \S 7.2, eq.~7-54; Gnedin, Hernquist \& Ostriker 1999).

If indeed $\bar\rho(\rperi) \sim \bar\sigma(\ellt)$, then 
the velocity kick integrated along a path of length $\sim \rperi$ near
pericentre is comparable to the typical particle velocity,
$\Delta v \sim v \sim \bar\rho(\rperi)^{1/2} \ellt$.  
This makes the energy change maximal in the outer
satellite regions, with comparable contributions from the linear and quadratic
terms. The kick inwards causes a delayed escape,
because the escapers need to cross the satellite 
before departure, and the crossing time is comparable to the
orbital period in the halo. Therefore, the actual stripping is expected to
occur in the halo near the following apocentre, 
where the particle is ``deposited" anyway
(\se{orbits}).

When apocentre is at $\alpha>1$ but pericentre is already in the $\alpha<1$ 
regime, the duration of maximum tides becomes longer and the impulse 
approximation less valid. Any stripping that still occurs along such an 
orbit ends up with ``deposit" outside the $\alpha<1$ core.

The shrinkage near pericentre and the following re-bounce and later mass
deposit near apocentre are clearly seen in the simulations of the radial and 
elongated mergers, \Fig{snaps} and \Fig{radii}.

Once the satellite orbit becomes confined to the inner halo where $\alpha < 1$,
we expect no impulse stripping.
If the slope near pericentre is $\alpha$, the tidal force drops to
half its peak magnitude at a distance of $\sim 2^{1/\alpha}\rperi$ 
from pericentre. At $\alpha=1/2$ this is already $4\rperi$,
so the tidal force is not much weaker than its pericentre value
throughout a large fraction of the orbit.  Furthermore, adiabatic
compression makes $\bar\sigma$ become larger than $\bar\rho(\rperi)$ 
(\se{ad} below). These two effects make the impulse approximation invalid.
The effect of the inwards tidal force accelerating an incoming particle is
roughly balanced by the inwards tidal force of similar magnitude 
working to decelerate the particle when it is going out.
In the limit where $\alpha=0$, the tidal force is the same in all
directions and constant in time, so $\int \bFt \d t=0$ along each particle 
orbit and there is no net impulse.

We conclude that we expect no impulse tidal mass deposit in the core region
where $\alpha < 1$: neither during the early quick core passages nor after 
the decay of the orbit into the core region. This is consistent with the
behavior in the radial and elongated merger simulations.


\subsection{The adiabatic limit}
\label{sec:ad}

The adiabatic invariance of actions, which is formally valid when the internal
orbital periods are much shorter than the characteristic
time over which the external potential is varying, is known to be crudely 
applicable even when these time scales are comparable to each other
(BT, \S3.6, Fig.~3-29).  Since $\bar\rho(\rperi) \sim \bar\sigma(\ellt)$,
and since for bound satellite particles along the satellite orbit
$\ell \leq \ellt$ and $r \geq \rperi$, 
The satellite particles inwards of the tidal radius
are likely to obey the adiabatic approximation at all times. 
Once the decayed orbit becomes confined to a central halo region
of $\alpha <1$, where the tidal force varies slowly along the orbit and
actually tends to a constant when $\alpha \rightarrow 0$,
the adiabatic approximation becomes crudely applicable also for the outer
satellite particles. Any adiabatic compression of the satellite would 
strengthen the applicability of this approximation.

In this case, the tides gradually distort the satellite, and may cause 
slow mass loss mostly outside a momentary tidal radius. 
Now tidal forces outwards cause stretching and energy gain which may lead 
to mass loss, while tidal forces inwards cause contraction, make the satellite
more bound and do not lead to stripping.

In order to gain some insight into the $\alpha$ dependence of adiabatic 
tidal effects we consider the following simple example.
We see in \equ{tide_harmonic} that the tidal perturbation in each direction
($i=1,3$) mimics the force of a harmonic oscillator about the satellite centre,
\be
\Fti = -\kti \ell_i
\,, \quad \kti \propto \bar\rho(r)\, [(1-\alpha),1,1] \,,
\ee
except that the force constant along $\bell_1$ is negative for $\alpha > 1$.
If we assume as a toy model that the satellite core is a homogeneous sphere
with density $\sigma_0$, 
the orbits under its own self gravity are also three-dimensional harmonic
oscillators, with
\be
F_{0i} = -k_{0i} \ell_i
\,, \quad k_{0i} \propto \sigma_0 \,.
\ee
This can be generalized to a homogeneous ellipsoid, with different force
constants in the different directions, based on Newton's third theorem
(e.g. BT, \S 2.3, Table 2-1).
The perturbed system is thus a 3-dimensional harmonic oscillator, with
force constants $k_i = k_{0i} + \kti$, and corresponding frequencies
$w_i^2 = k_i$.
If the motion of the satellite within the halo 
introduces slow variations in the tidal perturbation, and correspondingly
in $k_i$, the particles should conserve adiabatic invariants along their 
orbits.
The case of a radial satellite orbit within the halo is particularly simple
because the principle directions remain stationary.
The adiabatic invariant of the harmonic oscillator along each direction is
the action integral
\be
\int v_i^2 \d t \propto w_i L_i^2 \propto -E_i/w_i \simeq {\rm const.} \,,
\ee
where the integral is over the period of the oscillation, 
$L_i$ is the amplitude of the oscillation,
and $E_i$ is the corresponding energy (e.g, BT, \S 3.6.b).
Thus $\Delta E_i \propto -\Delta \omega_i$, meaning that
strengthening (weakening) of the force constant $k_i$, 
corresponding to an increase (decrease) in $w_i$, 
leads to shrinking (stretching) and negative (positive) energy changes.
This implies that escape may be caused only for particles moving
roughly along the radial-orbit line ($\pm \bell_1$)
and only as long as $k_1 \propto \bar\rho(r) [1-\alpha(r)]$ is decreasing while
$r$ is decreasing.  When the satellite is decaying within a halo core 
where $\alpha<1$ 
and where the forces are towards the satellite centre in all directions, 
all the force constants are clearly increasing with decreasing $r$, 
so there is systematic shrinking in all directions. The corresponding
energy changes are negative for all the particles, implying
no stripping.

When the satellite orbit becomes confined to a flat core of $\alpha=0$ and 
density $\rho_0$, the restoring force is constant and fully spheri-symmetric, 
${\bf F} \propto -(\rho_0 +\sigma_0) \bell$, independent of the actual
orbit of the satellite within the halo core.
The satellite is then in its most tightly bound configuration. 
We may expect the situation not to be very different for small $\alpha$ values
in the range $\alpha < 1$,
as the deviation from spherical symmetry is limited to one direction,
along which the force is still restoring and is getting stronger as $r$ 
decreases.

The general behavior in the merger simulations is consistent with the
behavior in the above toy model. 
We noticed in \Fig{radii} that
once the satellite becomes confined to the halo core
there is no further overall shell expansion, indicating
that stripping has stopped.
at $\alpha < 1$ can be qualitatively reproduced in the extreme limiting 
cases where the impulse or the adiabatic approximations may be crudely valid.
This provides some intuitive understanding for why
the slope of $\alpha \sim 1$ is a critical 
value below which mass transfer is inefficient.

\section{Tidal Radius}
\label{sec:tidal_radius}

We now turn to a complementary approach, where the satellite is addressed
as a spherical ``onion" from which the outer layers are gradually peeled 
off by the tides.
The $\alpha$ dependence of a slow stripping process can be estimated by
means of the momentary tidal radius.
This depends on the nature of the orbit of the satellite within the halo,
which dictates, for example, whether the most useful frame of reference for 
defining the tidal radius should be rotating or not.

A lower bound to the tidal radius is obtained rigorously 
in the (rare) case of a {\it circular} orbit.
If the circular angular velocity is ${\bf\Omega}$, 
the potential is stationary in a frame which co-rotates with the
same angular velocity,
in which the corresponding fictitious centrifugal 
force should also be taken into account.  In this case, the Jacobi integral $J$
is conserved along the orbit of each satellite particle (see BT, \S 3.3
and \S 7.3). This integral is the energy as measured in this rotating frame 
including in the effective potential the term corresponding to the centrifugal 
force, $-({\bf \Omega} \times \bx)^2$.
The outermost closed zero-velocity surface of constant $J$ defines a firm
closed boundary around the satellite, the Roche lobe, which cannot be crossed
by particles from the inside out.  Therefore,
the distance from the satellite to the Lagrange point, the saddle point of the
effective potential, provides a lower bound for the tidal radius.
This Lagrange point can be obtained by balancing in the rotating satellite 
frame the self-gravity of the satellite with the sum of the tidal and
centrifugal forces along the line connecting the centres of mass. 
The self-gravity pull inwards by the satellite mass interior
to $\ell$ is $-[Gm(\ell)/ \ell^3] \bell$
and the centrifugal force is of an opposite sign and the same amplitude
because of co-rotation [$\Omega^2 \propto M(r) / r^3 = m(\ell) /\ell^3$].
Using the $\bell_1$ component of the tidal force from \equ{tide_harmonic}
we obtain for the tidal radius when the satellite is on a circular orbit of
radius $r$:
\be
\alpha(r)\, \bar\rho(r) = \bar\sigma(\ellt)
\label{eq:tidal_circ}
\ee
(compare to BT, eq.~7-84, for the case $\alpha=3$).
In this case the combined effect of the tidal and centrifugal forces
along the line connecting the centres of mass is always pointing away from
the satellite centre, towards $+\hat{\bell}$.
This stretching force becomes weaker as $\alpha$ decreases, but it
completely vanishes only at $\alpha=0$.
In the direction perpendicular to the line connecting the centres of mass
but still in the merger plane the tidal compression
along $-\hat{\bell}$ is exactly balanced by the centrifugal force,
independent of $r$ and $\alpha$. In the direction perpendicular to the
merger plane, the centrifugal force is zero so the net force
is the usual tidal compression, with no explicit $\alpha$ dependence.

In the other extreme (more realistic) case of a merger on a {\it radial} orbit,
the only forces working on the
particle in the satellite (non-rotating) rest frame are the tidal force 
and the self-gravity pull inwards.
The direction of maximum tides is
stationary, and only the magnitude of the force varies as a function of $r$.
Even though there is no exact analog of the Jacoby integral in this case,
one may still define a tidal radius $\ellt$ by the 
point along the line of maximum tides at which the tidal and self-gravity
forces balance each other such that the net force in the satellite 
frame is zero. The condition replacing \equ{tidal_circ} for a radial orbit
is thus
\be
[\alpha(r)-1]\, \bar\rho(r) = \bar\sigma(\ellt) \,.
\label{eq:tidal_rad}
\ee

In the above expressions for the tidal radius, 
the satellite profile is crudely assumed to be fixed 
in time for $\ell < \ellt$ and its mass distribution is assumed to remain 
spherical; the shells outside the tidal radius are assumed
to be peeled like layers off an onion while the tidal distortions
are neglected.

We now investigate the evolution of the tidal radius, starting with
the radial-orbit case.
As $r$ is decreasing from the virial radius inwards, where $\alpha$ is large, 
the function $[\alpha(r)-1]\, \bar\rho(r)$ is increasing. Assuming that
$\bar\sigma(\ell)$ is a monotonically decreasing function, this means that
$\ellt$ is decreasing. More and more mass of the satellite is being torn
away as the satellites is moving in. However, as the satellite approaches
the position where $\alpha$ approaches unity, the function $[\alpha(r)-1]\,
\bar\rho(r)$ approaches zero, implying that the tidal radius grows to
infinity. This means that there is a halo radius $\rem$ and a corresponding 
slope $\alpham >1$ at which $[\alpha(r)-1]\, \bar\rho(r)$ obtains a maximum.
If the inner satellite density is high enough, then \equ{tidal_rad} implies 
that $\bar{\sigma}(\ellt)$ obtains a maximum value $\bar{\sigma}_{\rm m}$ 
at $\rem$, and therefore $\ellt$ obtains a minimum value $\elltm$ there. 
Since at $r<\rem$ (and $\alpha <\alpham$) the tidal radius derived by
\equ{tidal_rad} is larger than $\elltm$, we expect no further stripping inwards 
of $\rem$.  The slope $\alpham$ thus characterizes the point
of minimum tidal radius inside which the stripping stops,
and $\elltm$ marks the boundary of the central part of the satellite 
which remains intact, not to be disrupted.

The subsequent sinking in of this whole remnant to the halo centre due to
dynamical friction
causes a steepening of the profile at $r < \rem$ where $\alpha < \alpham$.
This steepening is effective at least as long as the inner slope
is flatter than $\alpha=1$.  It can be avoided only if the inner satellite
density is low enough such that $\ellt = 0$ already at $r>\rem$, namely
the satellite is fully disrupted outside the halo core region (see
\se{core} and \Fig{prof_puffy}).

As an example, assume that the {\it mean} halo density profile is given
by \equ{nfw} with $\rhoc=\rc=1$, namely $\bar\rho(r) = r^{-\ai}(1+r)^{\ai-3}$, 
and allow the inner slope $\ai$ to take a value between 0 and 3.\footnote{
The corresponding local density profile is a Hernquist profile,
$\rho(r) = (1-\ai/3) r^{-\ai} (1+r)^{\ai-4}$.}
The corresponding $\alpha$ profile is $\alpha(r)=(\ai+3r)/(1+r)$ 
and we find that $[\alpha(r)-1]\, \bar\rho(r)$ obtains a maximum at
$\rem =(1-\ai)/2 +(3\ai^2-12\ai+9)^{1/2}/6$.  
\Fig{alpham} shows, for a range of $\ai$ values, the corresponding
values of the minimum-tidal-radius quantities $\rem$, $\alpham$, 
and the corresponding $\bar\sigma(\ellt)$ according to \equ{tidal_rad}. 
We see that for a flat core, $\ai=0$, the minimum tidal radius is obtained
at $\alpham=1.5$ 
and $\rem=1$. For larger $\ai$ in the range $(0,1)$, the value of $\alpham$
gradually decreases towards $\alpham=1$, namely it converges to the asymptotic
value $\ai=1$ at $r \rightarrow 0$. For $\ai \geq 1$, the minimum tidal radius
is obtained at $\rem=0$, with $\alpham=\ai$. 
For $\ai$ in this range the stripping continues until the tidal radius 
shrinks to zero and the whole satellite is disrupted.

\begin{figure}
\vskip 6.3cm
{\includegraphics{f7.ps}}
\caption{
Quantities when the tidal radius obtains a minimum,
as a function of the halo inner slope $\ai$ in \equ{nfw}. Shown are
the position in the halo $\rem/\rs$, the slope $\alpham$
and the mean density within the minimum tidal radius  
$\bar{\sigma}_{\rm m}(\ellt)$. Note that $\alpham \geq 1$.
       }
\label{fig:alpham}
\end{figure}

An overall tidal compression at $\alpha<1$ should actually enhance the 
steepening effect.  It would compactify the satellite while in the halo core
and add halo mass to it, which would make the dynamical friction 
more effective and speed up the orbit decay to the halo centre.
Moreover, the halo would react to the addition of satellite mass to its centre 
by further contraction, preferentially at smaller radii, roughly obeying 
the adiabatic invariant that requires $M r \sim $const.
inside each shell of material.

In the case of a circular orbit, \equ{tidal_circ},
an analogous analysis to the one summarized in \Fig{alpham}, for the example
of a mean halo profile given by \equ{nfw}, yields for a core of $\ai=0$  
a minimum tidal radius at $\alpham=3/4$ (the analog of $\alpham=1.5$ in the
radial case). 
The minimum tidal radius becomes zero and $\alpham=\ai$ for any $\ai>3/13$
(the analog of $\ai>1$ in the radial case).  
Since \equ{tidal_circ} provides a lower bound to the tidal radius,
these values of $\alpham$ should be regarded as lower bounds for the
circular-orbit case. 

Since at a given position in a given halo central region (given $\alpha$) 
the stripping seems to be more efficient for satellites on circular orbits
than on radial orbits, and the compression is correspondingly less effective, 
the dynamical friction becomes weaker for circular orbits. 
This should slow down the dynamical-friction decay of the orbit to the 
halo centre and the steepening process, as noticed in our N-body simulations.

Thus, based on the current simplified analysis, the satellites on circular 
orbits are not necessarily as effective as those on radial orbits in 
steepening cores to cusps.
However, we argue in Dekel \etal (2003) (as summarized in \se{asymp} below)
that the $\alpha$ dependence of the mass-transfer efficiency is enough to
cause steepening of the profile even by satellites on circular orbits, 
as is demonstrated by the simulations (\se{sim2}).

The tidal radius of a satellite on a realistic, elongated orbit is likely
to lie in between the radii obeying \equ{tidal_rad} and \equ{tidal_circ}.
At large $r$ (compared to pericentre), an elongated orbit can be 
approximated as a radial orbit.  Near pericentre of an orbit about a 
point-mass halo the orbit might have resembled a circular orbit, 
but in a flat halo core the orbit near pericentre is better approximated
by a straight line.
Thus, the steepening effects of typical satellites on 
the inner halo profile are more likely to be similar to the effects 
evaluated for radial orbits.
Furthermore,
the fact that even in the simulation of a merger on a circular orbit 
the mass deposit
seems to practically stop near $\alpha \sim 1$, not far from the behavior 
in the radial-orbit case, indicates that the estimate of tidal radius based on
the Jacoby integral may indeed be an underestimate.

We conclude that the general behavior in the simulations
of tidal mass transfer as a function of $\alpha(r)$ can be qualitatively
understood also in terms of the idealized concept of stripping outside
a tidal radius.

\section{Convergence to an Asymptotic Cusp}
\label{sec:asymp}

The tidal mass transfer in halo regions where the slope is $\alpha>1$,
and the actual convergence from either side to an asymptotic cusp,
are analyzed in an associated paper (Dekel \etal 2003). 
We provide a quick preview of this analysis here for the completeness 
of the discussion.

We develop a simple prescription for tidal mass transfer by relating
every shell $\ell$ in the initial satellite with a ``deposit" radius $r$
in the halo. This is done by equating the encompassed mass within $\ell$
in the initial satellite with the mass of stripped satellite material
found inside halo radius $r$ after the merger, $m(\ell)=\mf(r)$.
This correspondence can be expressed in terms of $\alpha$:
\be
{\bar\rho(r) \over \bar\sigma(\ell)}
= \psi[\alpha(r)] \,,
\label{eq:strip_con}
\ee
where $\bar\rho(r)$ and $\bar\sigma(\ell)$ are as before
the initial mean density 
profiles of the halo and the satellite. What makes this prescription useful is
that $\psi(\alpha)$ is found to be quite insensitive to the specific nature 
of the merger (see below).
Several authors (e.g., Syer \& White 1998; Klypin 1999a) applied
the crude resonance condition $\psi=1$.
However, the $\alpha$ dependence of the tidal force, and in particular
its vanishing at low $\alpha$, imply that
$\psi(\alpha)$ should be a monotonically {\it decreasing} function of $\alpha$.
When we adopt the approximation of peeling layers off a 
fixed onion as above, we obtain from \equ{tidal_rad} and \equ{tidal_circ}
that $\psi \propto (\alpha-1)^{-1}$ and $\propto \alpha^{-1}$ respectively.
When we also take into account the 
deposit at apocentre due to stripping at pericentre
and the stretching and distortion of the inner 
satellite mass before it is being torn away\footnote{a 
decrease in satellite density before stripping
has been seen in simulations (e.g. Klypin \etal 1999a, Fig.~6; Hayashi \etal
2002)}, 
we expect $\psi$ to be smaller than unity
even near $\alpha \sim 1$ and to decline further with $\alpha$ at $\alpha>1$.

We then measure $\ell(r)$ and obtain $\psi(\alpha)$
in several different merger simulations like the ones described above.
The measured $\psi(\alpha)$ is found to be qualitatively similar 
for the different merger cases, despite the very different orbits
and initial density contrasts between satellite and halo.
[This similarity can be seen
from the resemblance of the final profiles $r(\ell)$ in \Fig{radii}].
An approximate empirical fit is obtained, for example, 
by $\psi(\alpha)=0.5/\alpha$, with a spread of less than $\pm 0.1$ about it 
in the different cases studied.
The robustness of $\psi(\alpha)$ indicates that \equ{strip_con} can serve 
as a useful approximate recipe for tidal mass transfer in the general case.
We conclude that the simplified resonance condition, $\psi=1$, does not provide
a good approximation for where the satellite mass ends up.
The actual mass transfer is more efficient ($\psi <0.5$), and its relative
efficiency gets higher in steeper regions of the halo profile.

\begin{figure}
\vskip 6.5 truecm
\includegraphics{f8.ps}
\caption{
A schematic illustration of satellite mass deposit in the halo.
Shown are an NFW halo profile $\bar\rho(r)$, and a homologous satellite
$\bar\sigma(\ell)$ properly shifted to the left and upwards.
The arrows connect shell radii $\ell$ to the halo radii where they are
deposited $r$.
The horizontal dashed arrows refer to stripping when stretching is ignored,
$\psi(\alpha)=1$. This would steepen the profile, as steep regions of
$\bar\sigma(\ell)$ are deposited at flatter regions of $\bar\rho(r)$.
The solid arrows illustrate realistic stripping after stretching.
The vertical displacements, which grow with $r$, refer to
$\psi(\alpha)<1$.
The slope at $\ell$ may be flatter than the slope at $r$ such that the
mass tends to be deposited at larger $r$ and the result is flattening
of the halo profile.
}
\label{fig:nfw_hs}
\end{figure}

If the mass transfer is described by \equ{strip_con} 
with $\psi(\alpha)$ declining rapidly enough,
we show in Dekel \etal (2003) that
the profile evolves slowly towards an asymptotic
stable power law with $\aas \gsim 1$.
We assume that the halo and satellite are drawn from a cosmological
distribution (as predicted theoretically and seen in simulations); 
they are homologous,
with their characteristic radii and densities scaling like
$\ellc/\rc \propto m^{(1+\mnu)/3}$
and
$\sigc/\rhoc \propto  m^{-\mnu}$,
where $\mnu \simeq 0.33$ for $\Lambda$CDM.
\Fig{nfw_hs} helps understanding intuitively the origin of an asymptotic 
slope due to the decreasing nature of $\psi(\alpha)$.
We write
$\bar\rho_{\rm final}(r) = \bar\rho(r) +\bar\sigma(\ell) {\ell^3/ r^3}$,
and obtain for the change of the local slope in a merger
\be
\Delta \alpha(r)
\propto  -{\dd \over \dd r}
\left( {\bar\sigma(\ell) \over \bar\rho(r)} {\ell^3 \over r^3} \right)
\, .
\label{eq:dalpha} 
\ee
While every power law is a self-similar solution,
$\Delta\alpha(r)=0$, it is not necessarily a stable one.
For example, with $\psi=$const.~one would have obtained $\Delta\alpha(r)>0$ 
everywhere (because $\ell/r$ is decreasing with $r$),
namely a continuous steepening towards $\alpha=3$. 
On the other hand, a realistic mass transfer where  
$\psi$ is decreasing with $r$ may produce a stable fixed point 
where $\Delta\alpha=0$ and the second derivative is negative.
A rigorous linear perturbation analysis determines the rate of
convergence and yields an equation for the value of the asymptotic slope
$\aas$ under a sequence of mergers with the same mass ratio $m/M$:
$$
\Delta\alpha \,\propto\, \alpha(\alpha - 3){\psi'(\alpha)}/{\psi(\alpha)}
+ 3 \ln[(m/M)^{-\mnu} \psi(\alpha)]  = 0 \,.
\label{eq:aas}
$$
The solutions for $m/M$ large enough are typically in the range 
$1 < \aas \leq 1.5$.
For a sequence of mergers with a cosmological distribution of mass ratios
we obtain an asymptotic slope comparable to the solution of \equ{aas}
with $m/M\simeq 0.3$.

\begin{figure}
\vskip 5.1truecm
{\includegraphics{f9.ps}}
\caption{
Toy-simulation evolution of slope $\alpha$ at $r=0.1\rc$
due to a sequence of mergers with mass ratio $m/M=0.3$.
The initial profile is \equ{nfw} with $\ai$ either zero or 2.
When $\alpha<1$, the slope steepens rapidly to $\alpha>1$ within a few
mergers and then it converges slowly from either side towards an asymptotic
value.
}
\label{fig:aas}
\end{figure}

\begin{figure*}
\vskip 6.7cm
{\includegraphics{f10a.ps}}
{\includegraphics{f10b.ps}}
\caption{
Halo density profile before (solid) and after (dashed) the merger
with a puffy satellite, whose density at the characteristic radius
has been scaled down by a factor of about 2.5 
compared to the compact satellite.
Left: the radial merger simulation.
Right: the circular merger simulation.
Notation is as in \Fig{prof_compact}.
The puffy satellite loses most of its mass in the outer halo, leaving
the
halo core practically unaffected.
}
\label{fig:prof_puffy}
\end{figure*}

In order to test the linear analysis, we perform
toy simulations of the profile buildup by cosmological mergers,
where we implement the mass-transfer recipe, \equ{strip_con}, 
with $\psi(\alpha)=0.5/\alpha$.
\Fig{aas} shows the convergence of $\alpha$ at a fixed $r$ to the 
asymptotic value.  The profile actually evolves through momentary profiles
which are more relevant for comparison with real haloes at
different times during their buildup process; they resemble in shape the
generalized NFW profile, with an inner cusp growing slowly from $\alpha=1$ to 
$\alpha=\aas$. We find a weak sensitivity to the cosmological 
power spectrum of perturbations, partly because of the 
robust tidal effect driving the slope to $\alpha >1$, as
discussed in the current paper.

\section{A Flat Core by Puffy Satellites}
\label{sec:core}

Our robust result so far is that a cusp of $\alpha>1$ develops
in dark-matter haloes whenever mergers with compact satellites
bring enough bound clumps to reside 
in the inner parts of haloes. It implies that a necessary
condition for maintaining a flat
core is preventing bound clumps from settling in the core. This can be 
achieved within the $\Lambda$CDM scenario only if the cores of satellite 
haloes were puffed up significantly by 
processes other than the gravity of the cold dark matter. 
In this case the satellites would practically disrupt while the
apocentre of their orbit is still in the outer halo, and the halo core
could remain undamaged.

How much puffing is needed?
In order to obtain a first clue, we have performed simulations 
similar to those described in \se{sim1} and \se{sim2},
except that the compact satellites of CDM have been replaced by more puffy
satellites of the same total mass. 
In the Hernquist profile, \equ{sat_prof}, where the default compact satellites
had $\sigc=19.2$ and $\ellc=1$, the parameters of
the puffy satellites are now $\sigc'=\sigc/8$ and $\ellc'=2\ellc$. 
With this choice, the density at $\ell=1/2$, the characteristic radius
of the original compact satellite, is scaled down by a factor of $2.3$,
to just below the density of the host halo at its characteristic radius.
This corresponds to a reduction by a factor of $2.8$ in the
mean density interior to $\ell=1/2$.
\Fig{prof_puffy} shows the effect of such a merger with a puffy satellite
on the halo density profile in the case of circular and radial mergers.
When the satellite is puffy, we see that
almost all the satellite mass is stripped before the satellite
orbit decays to the $\alpha \leq 1$ zone, and as a result the halo core
is practically unaffected. We learn that a modest reduction in the 
initial satellite inner density is enough for preventing the cusp 
formation seen in \Fig{prof_compact}.
%

In fact, mergers with puffy satellites is a plausible mechanism for
turning a halo cusp into a flat core. A sequence of such mergers, in which
the stripped satellite mass is added to the halo outside its inner region,
may lead to a continuous increase of density in the outer halo while
the inner density remains unchanged, and thus flatten the inner profile.
This process is a subject of ongoing work beyond the scope of the
present paper; at the moment it is only a sensible speculation.
Our solid result so far is limited to the fact that puffing of the 
satellites is a necessary condition for halo core survival in the CDM
scenario.

Could baryonic feedback effects be responsible for the puffing up of
haloes that could help the haloes maintain their cores, or even
turn halo cusps into cores?
It is known, for example, that the dissipative infall of the baryons into the
halo centre causes adiabatic contraction of the inner dark-matter halo.
If a significant fraction of the baryons is then blown out of the galaxy 
with a velocity much higher than the escape velocity (e.g., as observed
by Adelberger \etal 2002), then
the inner halo may expand in response to a configuration more extended
than the original configuration before baryonic infall.\footnote{Note,
for example, that a virial system that instantaneously loses 
one half of its mass becomes totally unbound.}
It has been estimated that even in the extreme limit of instantaneous blowout
of all the gas the effect on the halo is not strong enough for
straightforwardly turning a cusp into a core in a final large halo
(Geyer \& Burkert 2001; Gnedin \& Zhao 2002; though 
Navarro, Eke \& Frenk 1996 find a stronger effect). 
However, this effect is possibly sufficient for the necessary indirect
puffing-up of the merging small haloes.
Gnedin \& Zhao (2002) estimate that direct feedback effects may
reduce the central halo densities by a factor of 2 to 6. Based on our
simulations with puffed-up satellites, this may be enough by
itself to avoid the steepening from a core to a cusp in the framework of the 
merging scenario. 
Furthermore,
the strength of the effect depends on how deep in the potential
well the gaseous disk resides before it is blown away, 
which is determined by the baryonic spin.  
The possibility that baryons in the merging satellites have lost angular 
momentum due to over-cooling before regaining it by feedback 
(Navarro \& Steinmetz 2000 and references therein; Maller \& Dekel 2002, see
\se{conc} below)
allows one to consider lower spin values than measured in today's
disk galaxies, and thus obtain stronger effects on the dark matter.

In fact, one can think of special circumstances within the hierarchical
buildup scenario that may boost up the feedback effects even further
and help maintaining flat cores.
We sketch here such a scenario, 
based on enhanced feedback in merging satellites due 
to the tidal compression of the baryons.
Consider a satellite made of dark-matter and a typical fraction of
baryons merging
with the halo on a typical eccentric orbit, which takes the satellite 
through the halo core to an apocentre in the outer halo a few times before it 
decays to inside the core region (as in our radial and elongated merger 
simulations).
Assume that cooling in the satellite is efficient such that before the
satellite passes through the halo core the baryons are already concentrated 
in the satellite core, making, say, a half-and-half mixture with the 
dark matter there.
In every passage of the satellite through the halo core, where $\alpha < 1$, 
the tides compress the satellite into high densities (\Fig{radii}), 
creating shocks and stimulating an efficient burst of star formation. 
(The tides may also induce accretion of material 
including massive stars from the core of the host galaxy onto the satellite).
By the time the satellite is turning around in the following apocentre
outside the halo core, the massive stars have produced supernova-driven winds
which drive much of the gas out of the satellite. 
The satellite have already lost much of its outer dark-matter envelope at 
this point and is basically made 
of the original satellite core plus most of the satellite baryons.
If the satellite core loses half its bound mass in this gas blow-out,
and the gas expulsion is instantaneous, then the satellite becomes completely
unbound. But even if the expulsion is slow, the adiabatic invariants 
imply a density drop by a factor of $\sim 8$.
Thus, the remaining satellite is now much more susceptible to tidal stripping,
which could disrupt it completely before it manages to settle in the halo core.
This scenario should be investigated in detail before one can consider it
viable. At this point it is just an intriguing speculative proposal for
enhanced feedback effects, based 
on the wisdom gained so far regarding the tidal effects, and very crude 
common wisdom regarding star formation and feedback (e.g. Dekel \& Silk 1986).

\section{Discussion}
\label{sec:conc}

The main result of this paper is that
tidal compression enforces an inner cusp of slope $\alpha > 1$ in dark-matter
haloes which are subject to tandem mergers with relatively compact satellites
as in the $\Lambda$CDM cosmology. The lower bound of $\alpha \simeq 1$
coincides with the transition from tidal forces outwards (at $\alpha >1$)
to tidal forces inwards (at $\alpha <1$), resulting in no local mass transfer
from the satellite to the halo near its core boundary and therefore 
steepening of the core into a cusp. 
This robust effect is one way to 
explain the halo inner structure as seen in cosmological N-body 
simulations of the hierarchical clustering scenario. It then helps 
providing a tool for addressing other processes 
which may explain the observed flat cores in (at least some) galaxies.
More details of the cusp formation are provided in an associated paper
(Dekel \etal 2003), where we derive a simple prescription for tidal mass 
transfer between satellite and halo,
and use it to show that successive cosmological mergers 
slowly lead to a stable asymptotic cusp that is only slightly steeper than 
$\rho \propto r^{-1}$.

The buildup of a cusp by a single merger with a compact satellite,
and in particular the dependence of the tidal mass transfer on the slope 
of the halo density profile, have been 
demonstrated using controlled N-body simulations of single isolated mergers.
The simulations also indicate that the cusp remains stable under successive,
similar mergers.
We tried to interpret in simple qualitative terms the detected
phenomenon of little mass transfer at $\alpha \lsim 1$, which gives
rise to the steepening to a cusp.
For this purpose we used several idealized toy-model considerations 
based on standard approximations, each valid in another limit 
and all pointing in the same direction.
It is evident that these are simplified approximations, and we expect
each to be crudely valid at most during part of the time and for 
limited parts of the satellite. Nevertheless, we recall that these
approximations sometime provide useful qualitative estimates even where 
there is no formal justification for them to be strictly valid.
The fact that in these different regimes the different 
approximations all point to the phenomenon we see in the simulations 
may be interpreted as a hint for a qualitative understanding
of this phenomenon.
One should not use these approximations for making accurate quantitative
predictions away from their formal range of validity, and our toy-model 
analysis is not meant to replace a detailed and definitive exploration 
using exact analytic dynamical techniques, which should be
the target of subsequent work. 
However, at this point the cosmological N-body simulations do not give us a 
clue for the actual origin of the cusp, and the exact methods have
not provided a satisfactory understanding yet either.
Given that, the controlled merger simulations which highlight a specific
key effect, and the idealized approximations which provide hints in the
same direction, are useful for making progress in the understanding of
the phenomenon associated with the merger process, 
and in particular for putting forward for further 
investigation a possible mechanism for cusp formation. 

Syer \& White (1998) (and independently Nusser \& Sheth 1999 and Subramanian,
Cen \& Ostriker 2000) also addressed the profile resulting from mergers.
Although they all see the formation of a cusp in many cases, our results show
several differences which one should try to understand. 
Syer \& White implemented the more simplified model of tidal stripping at 
resonance, $\psi=1$ in our terms, 
ignoring the $\alpha$ dependence (and in particular the compression
at $\alpha <1$ and the gradual stretching at $\alpha >1$).
According to their algorithm, the steepening or flattening of the profile
is determined solely by whether the satellite inner density is higher or lower
than that of the halo. 
They argue that their prescription allows a long-term survival of a relatively
flat core of $\alpha <1$ and it introduces a strong explicit dependence on 
the power spectrum of fluctuations
--- both in conflict with the findings in cosmological simulations.  
For each of several different power spectra they followed a 
sequence of mergers and saw an apparent convergence to a different
self-similar profile.
When we substitute the oversimplified recipe 
$\psi=1$ in \equ{dalpha}, we find that 
$\Delta\alpha$ is always positive, for any merger and at any $r$.
Indeed, when trying to repeat the Syer \& White toy simulations
with higher resolution and following more mergers we actually find that
the profile does not really converge to a stable cusp but rather
continue to steepen slowly towards $\ai=3$. 
Only when using the improved stripping prescription
where $\psi$ is decreasing with $\alpha$
do we obtain convergence to a flatter asymptotic profile. 
Furthermore, the asymptotic profile obtained via the revised stripping recipe
is more robust to the cosmological model. This is partly because 
the effect of tidal compression drives the profile to $\alpha \gsim 1$
independently of the fluctuation power spectrum, and because 
the cosmological dependence of the ultimate asymptotic slope is weakened 
by the fact that the profile is determined by mergers of relatively
large mass ratio (see Dekel \etal 2003).
 
Somewhat puzzling is the finding of power-spectrum dependence in the profile
resulting from a sequence of N-body merger simulations by Syer \& White,
and especially the apparent slope flatter than $\alpha=1$ found 
in one of their cases.
It is puzzling because these simulations should have automatically included the 
correct tidal effects.
We can only suspect that the main shortcomings of their N-body
simulations are the limited resolution within the cusp region (only 8000
particles in the whole halo)
and perhaps the fixed narrow range of satellite masses.

The gravitational processes leading to a cusp are modeled in our analysis
as tidal effects during the buildup of the halo by a sequence of {\it mergers}.
This picture is likely to be valid in the CDM hierarchical clustering scenario,
where numerous sub-galactic haloes exist on all scales and are continuously 
merging (e.g., Klypin \etal 1999b; Moore \etal 1999a; Springel \etal 2001).
This merger picture
is confirmed by a careful inspection of high resolution CDM simulations,
where a special effort is made to identify the merging clumps which otherwise
could have been easily missed (e.g., Wechsler,
Dekel \etal~, in preparation).
It is therefore clear that cores, independently of how they form,
cannot survive in a pure CDM scenario without some modification --- they
efficiently turn into cusps as a result of the mergers with the relatively
compact building blocks.
 
Nevertheless, a cusp, though somewhat flatter, is reported to be seen also in 
simulations where the 
initial fluctuations had less power on small scales, thus suppressing the 
number of sub-galactic satellites and the associated merger rate
(Moore \etal 1999b; Avila-Reese \etal 2001; Bullock, Kravtsov \& Colin 2002).
Further indications for the generality of cusp formation comes
from simulations by Huss, Jain \& Steinmetz (1999), who find that cusps
also form as a result of collapse from roughly spherical initial density
perturbations with random velocity perturbations, as well as from the
simulations of Alvarez, Shapiro \& Martel (2002), who find that
cusps arise from the gravitational instability and fragmentation
of cosmological pancakes.
One way to explain this is by noticing that 
the asymptotic cusp formed in the CDM scenario is driven by mergers 
with relatively massive satellites (of typical mass ratio 1:3, 
see Dekel \etal 2003), and realizing that such mergers do happen even 
when small-scale power is suppressed and when pancakes fragment. 
It is therefore possible that the cusp is actually driven by mergers 
to a certain extent
also in these cases.
If it turns out that a cusp also forms when the merger picture is strictly 
invalid, it would imply that the gravitational processes 
involved in the halo buildup somehow mimic a behavior similar to the 
merger case.  We note in particular that the tidal compression discussed
above is expected to amplify density perturbations and possibly make them 
behave in certain ways like merging satellites. 

More generally speaking, however, one should accept the fact
the halo buildup is a complex gravitational process, whose different aspects
can probably be modeled in more than one way, 
e.g., as a violent relaxation process driven by fluctuations in comparison
with a sequence of mergers and substructure accretion.\footnote{We 
encounter an analogous duality, for example, when one manages to explain 
the origin of galactic spin alternatively via tidal torque theory applied 
to shells and by summing up the orbital angular momenta in a cosmological 
sequence of mergers (see Maller, Dekel \& Somerville 2002).}
There is also the intriguing possibility that general statistical
considerations in phase space may provide some clue for the origin of
a universal halo profile (e.g., Taylor \& Navarro 2001). 
Unfortunately, we currently know of no viable alternative to the tidal effects
in mergers as a simple model for
the origin of $\alpha \gsim 1$ cusps in the cosmological simulations.
The idealized merger picture provides one possible toy model, 
properly valid at least in hierarchical clustering scenarios,
within which the origin of the cusp is understood in 
simple terms.

Our result implies a {\it necessary} 
condition for the survival of cores in haloes independent of their origin
--- that satellites should be prevented from adding mass to the halo cores.  
This could be avoided in a CDM scenario if feedback processes manage to puff up
the small haloes and make them disrupt before they merge with the halo cores.
We have not explicitly addressed in this paper the {\it sufficient}
conditions for 
the formation of cores, but one can imagine that a sequence of mergers with
low-density satellites, where the mass is predominantly deposited 
outside the inner halo region, would indeed flatten the inner profiles
(work in progress).

Supernova feedback effects are probably not strong enough for turning
cusps into cores in haloes with rotation velocities higher than $\sim 100\kms$,
but the feedback may be sufficient for lowering the inner densities in the 
small halo progenitors such that when they merge they give rise to cores.
However, if the (so far inconclusive) observational clues for cores in  
clusters of galaxies are confirmed, simple supernova 
feedback is unlikely to provide a viable explanation for their origin.
In this case one may search for higher efficiency in supernova feedback
either due to microscopic effects such as porosity
in a muliphase ISM or due to hypernova from very massive stars (Silk 2002).
Alternatively, one may appeal to stronger feedback mechanisms, perhaps
associated with radio jets from AGNs. This process may be indicated
by an observed correlation between AGN activity and bright galaxies
in SDSS (Kauffmann \etal 2003, in preparation), 
and it may be needed independently in order to explain the missing baryons 
in big galaxies (Klypin, Zhao \& Somerville 2002)
and in clusters.
Otherwise, 
such large cores may present a real challenge for the standard 
hierarchical clustering scenario.

Other processes may also contribute to the development of halo cores.
One proposed scenario is based on scattering of inner halo particles off 
central, massive black holes (e.g., Merritt \& Cruz 2001). This could
be useful in explaining the centres of early type galaxies, but one might
suspect that it would be difficult to apply this scenario on the  
scales of $\sim 10$kpc corresponding to halo cores, which are
much larger than the scales associated with the black hole(s).
A second scenario addresses the heating of halo particles
by gas clouds spiraling in due to dynamical friction
(El-Zant, Shlosman \& Hoffman 2002). This is based on assumed specific
gradients of velocity dispersion in the inner haloes, which may or may not
be valid.
A third explanation is based on an efficient angular-momentum transfer
from a big rotating bar into the halo (Weinberg \& Katz 2002)./

However, other simulations indicate that this effect is inefficient
(Valenzuela \& Klypin 2002).  
It has been demonstrated (Sellwood 2002)
that the Weinberg \& Katz result is likely to be an artifact of
the unrealistically large bar assumed, and the representation of this
bar by a rigid body.
We stress again that, no matter what the origin of the core might be, 
our analysis implies that such cores could survive only if 
they are not perturbed by significant mass transfer from merging 
compact satellites, which implies that small CDM haloes must be puffed up
before they merge into bigger haloes. We also stress that none of these
scenarios seem to be capable of explaining the formation of cores in 
haloes on the scales of clusters of galaxies within the CDM scenario.

A possible caveat, which could in principle jeopardize our results as well as
the whole current standard paradigm regarding the validity of N-body
simulations in demonstrating robust cusp formation, 
has been raised by Katz \& Weinberg (2002).
They claim that all current cosmological simulations fail to recover 
the actual formation of a core in dark-matter haloes because the N-body 
noise smears out a delicate resonant reaction of halo-core orbits to the tidal
perturbation by the satellite. Based on linear analysis and smoothed-potential
simulations, they argue that an induced global mode
dominates the process and invalidates the standard dynamical friction
approximation by Chandrasekhar. As a result, satellites do not
decay into the core but are rather disrupted outside it, thus maintaining
a flat core profile at small radii.  
The dominance
of this effect in realistic cases has not been conclusively demonstrated.
It is hard to imagine how the resonances can remain so isolated and stable
in systems where phase mixing and broadening of resonances are expected.
The linear analysis is far from being valid in the nonlinear
case in hand, and the over-smoothing in their simulations may hide
physical gradients that may spread out the resonances (A. Klypin, private
communication). 
Even if this resonant behavior turns out to be important in the 
$N\rightarrow\infty$ limit, its relevance in real galaxies is questionable,
where the baryons are likely to 
introduce clumpiness at a level comparable to the granularity in 
current N-body simulations.  Still, this is a caveat worth pursuing.

The cusp/core problem is only one of the difficulties facing galaxy formation
theory within the CDM cosmology. It turns out that other main
problems can also be modeled by tidal effects in mergers, and may also be 
resolved by the inevitable feedback processes. For example,
Maller \& Dekel (2002) addressed the
angular-momentum problem, where simulations including gas
produce disks smaller than the galactic disks observed
(Navarro \& Steinmetz 2000 and references therein; Governato \etal 2002),
and with a different internal distribution of angular momentum
(Bullock \etal 2001b; van den Bosch, Burkert \& Swaters 2001).
A toy model has been constructed for the angular-momentum buildup by mergers
based on tidal stripping and dynamical friction, which helps us understand
the origin of the spin problem as a result of over-cooling in satellites.
A simple model of feedback has then been incorporated, 
motivated by Dekel \& Silk
(1986). This model can remedy the discrepancies, and in particular
explain simultaneously the low baryon fraction and angular-momentum 
profiles in dwarf disk galaxies.
 
Various feedback effects may also provide the cure to the missing dwarf 
problem,
where the predicted large number of dwarf haloes in CDM can possibly match
the observed number of dwarf galaxies only if the mass-to-light ratio
in these objects is very high (Klypin \etal 1999b; Moore \etal 1999a;
Springel \etal 2001; Kochanek 2001).  
Bullock, Kravtsov \& Weinberg (2000), Somerville (2002) and 
Tully \etal (2002)
appeal to radiative feedback effects which prevent the formation of
small dwarfs after cosmological reionization at $z\sim 7$, 
Scannapieco, Ferrara \& Broadhurst (2000) and Scannapieco \& Broadhurst (2001)
address the destructive effect of outflows from one galaxy 
on neighboring protogalaxies via ram pressure,
and Dekel \& Woo (2002) study the role of supernova feedback in determining
the relevant global properties of dwarfs and larger low-surface-brightness 
galaxies.
We note that
while the requirements from feedback in explaining the dwarf-galaxy 
properties and the angular-momentum problem are not too demanding, the 
solution to the core problem requires that the dark-matter distribution
be affected by feedback, which is a non-trivial requirement.
 
Nevertheless,
the successes of such toy models in matching several independent observations
indicate that they indeed capture the relevant basic elements of the complex
processes involved, and in particular that feedback effects may indeed
provide the cure to some or all the main problems of galaxy formation theory
within the $\Lambda$CDM cosmology that does so well on larger scales.
The alternative solution involving Warm Dark Matter (e.g., Hogan \& Dalcanton
2000; Avila-Reese \etal 2001; Bode, Ostriker \& Turok 2001) 
seems to still suffer to some extent from the core problem, 
it may still fail
to reproduce the angular-momentum profile in galaxies
(Bullock, Kravtsov \& Colin 2002), and it may be an 
overkill where the formation of dwarf galaxies is totally suppressed once the 
inevitable feedback effects are included (Bullock 2001).
The speculative alternatives involving self-interacting dark matter 
are even more problematic 
(Spergel \& Steinhardt 2000; Dave \etal 2001; Hennawi \& Ostriker 2002).

\section*{Acknowledgments}
We acknowledge stimulating discussions with Itai Arad, 
George Blumenthal, Andi Burkert, Doug Lin and Gary Mamon.
This research has been supported
by the Israel Science Foundation grant 213/02,
by the US-Israel Bi-National Science Foundation grant 98-00217,
by the German-Israel Science Foundation grant I-629-62.14/1999,
and by NASA ATP grant NAG5-8218.


\label{lastpage}
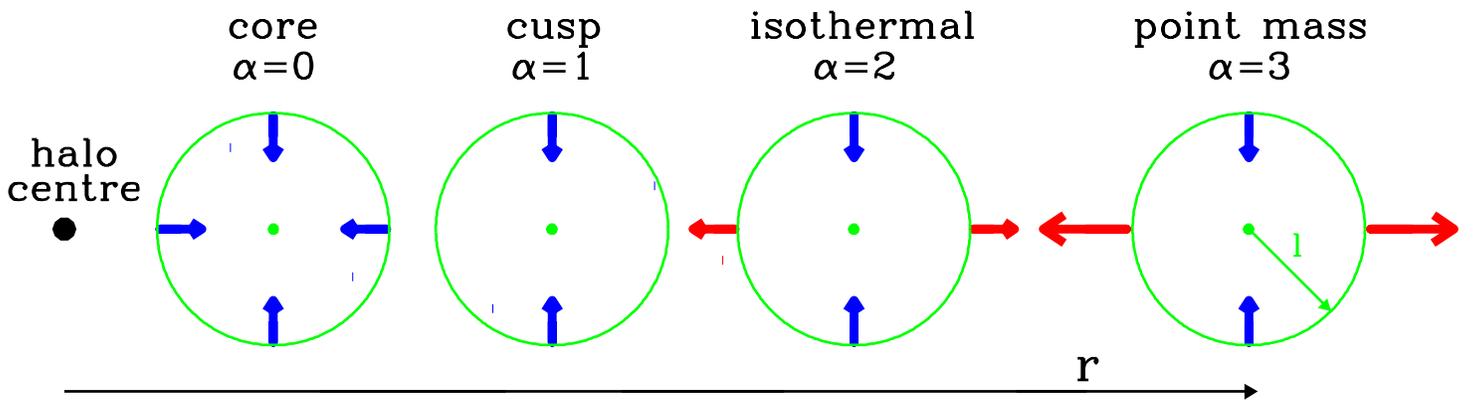
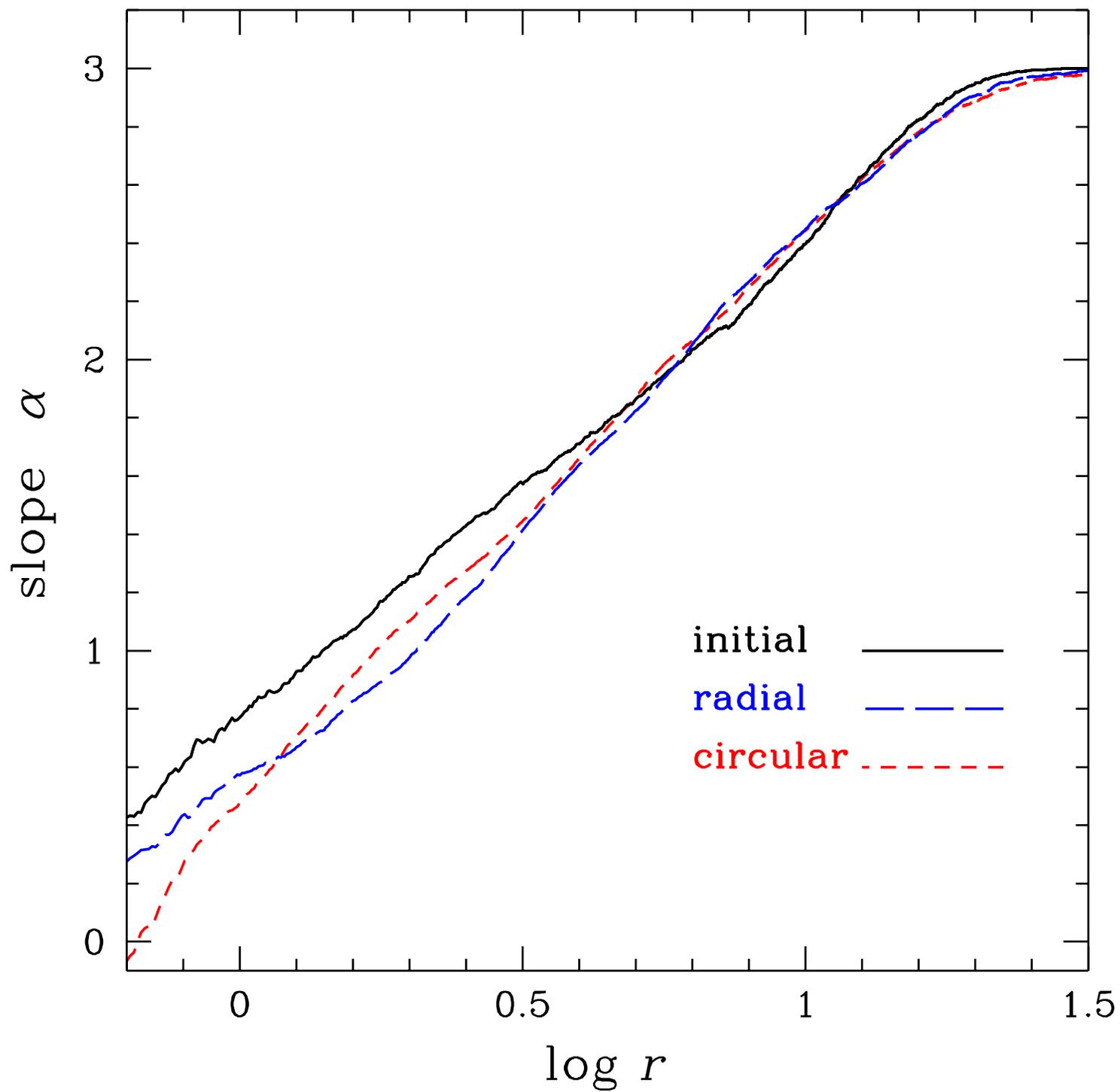
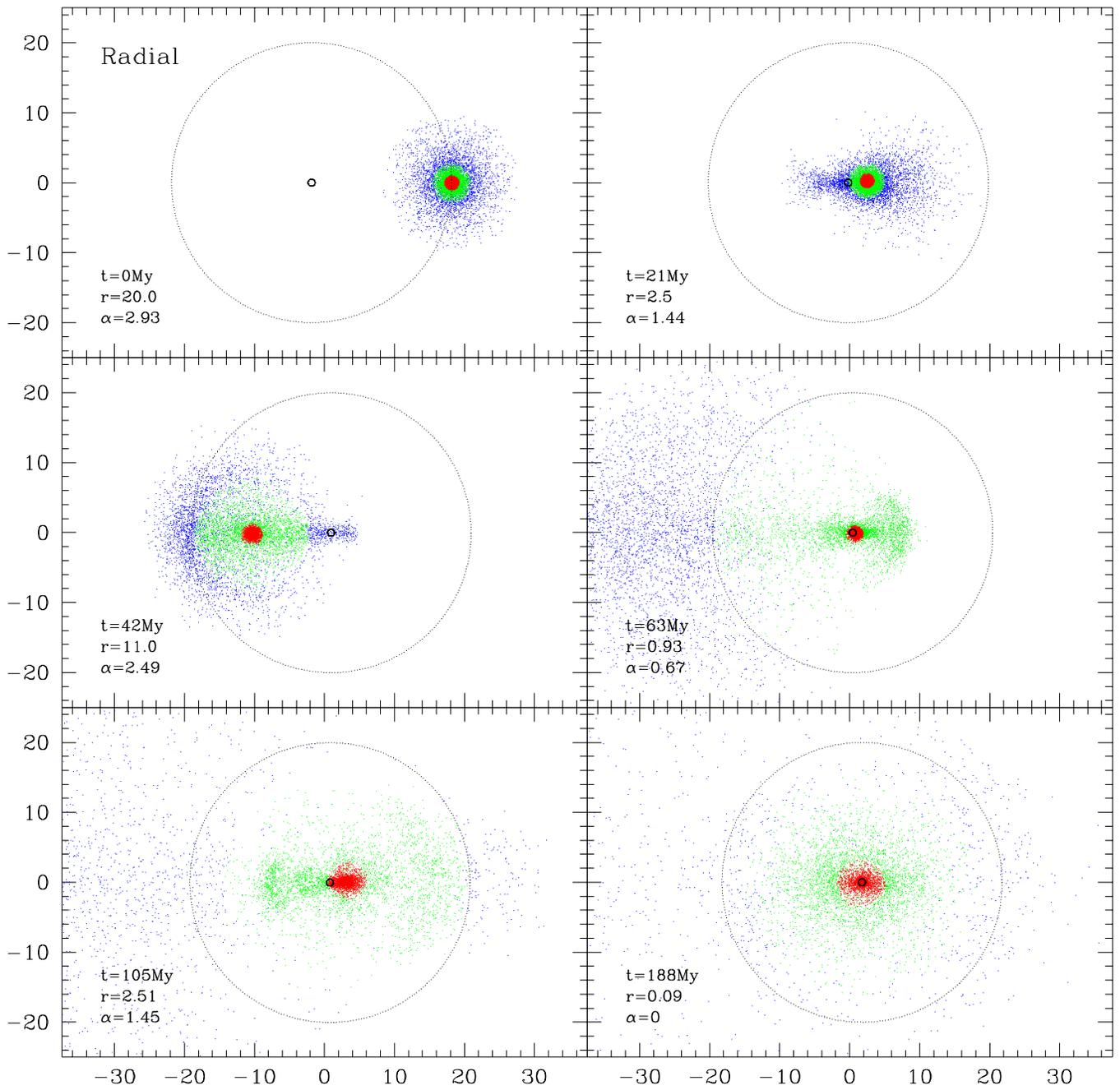
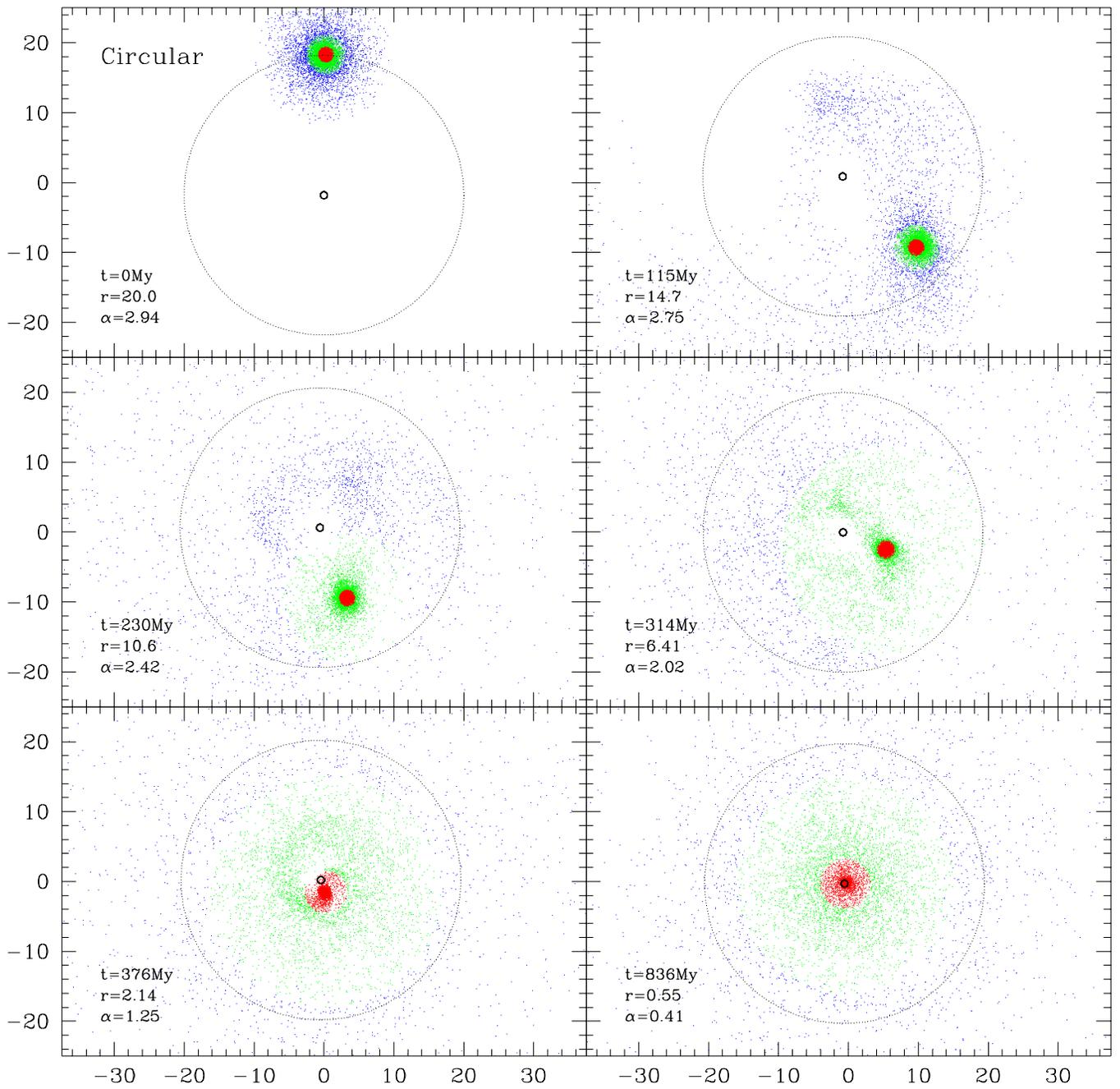
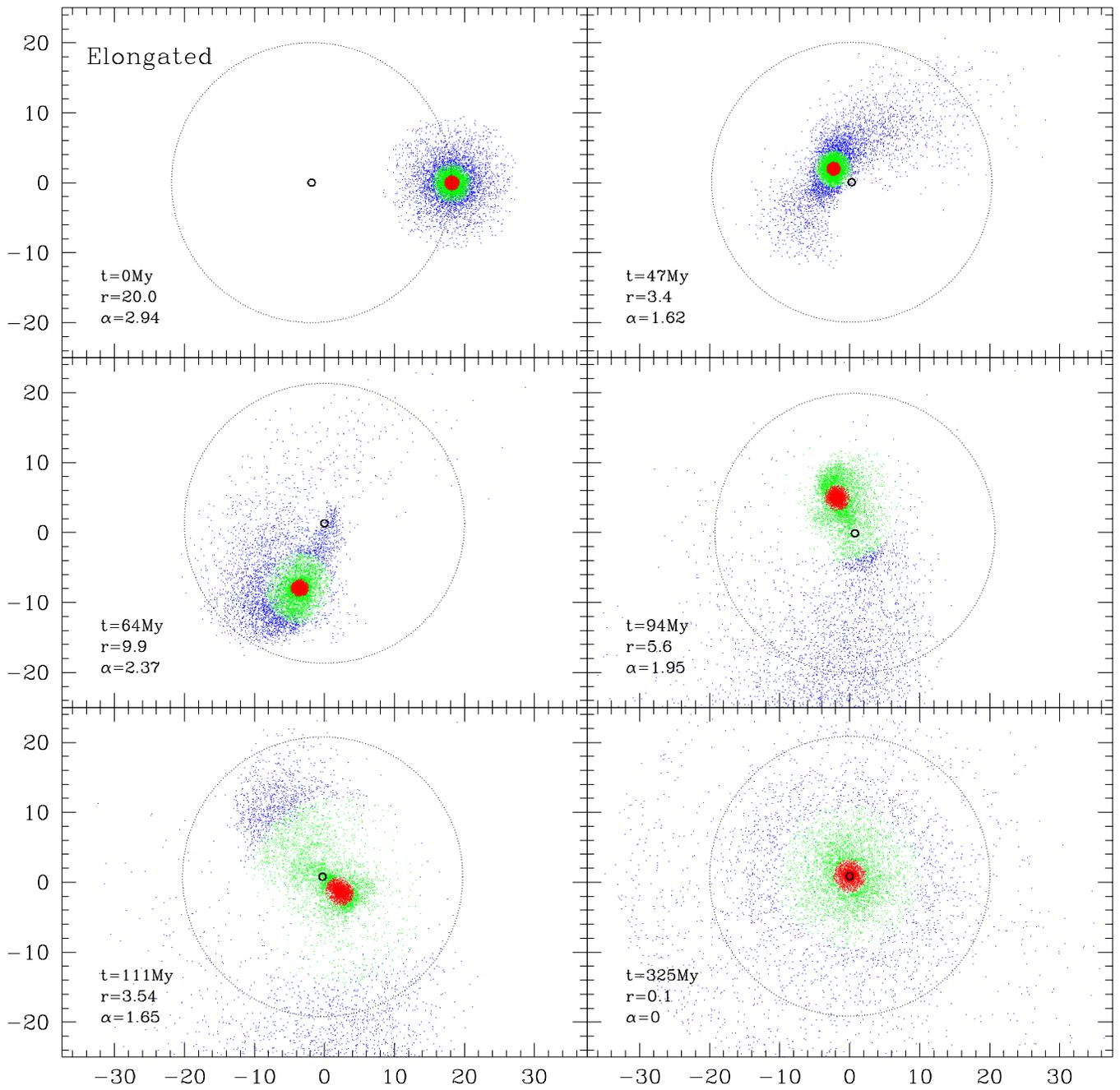
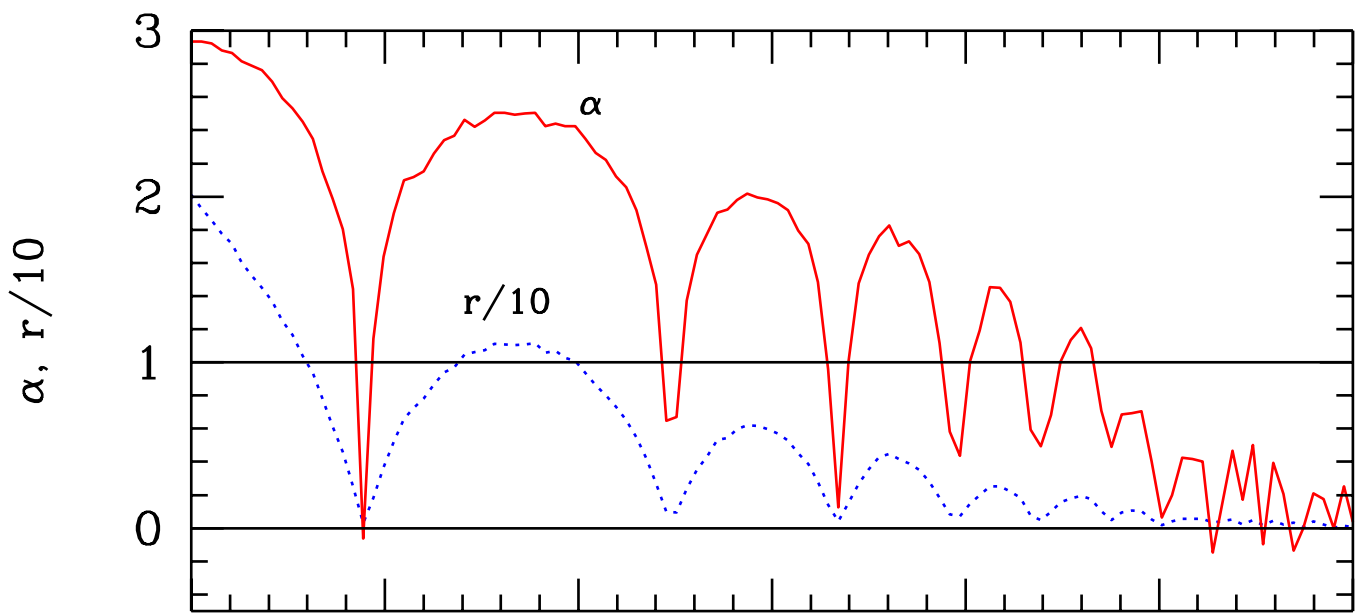
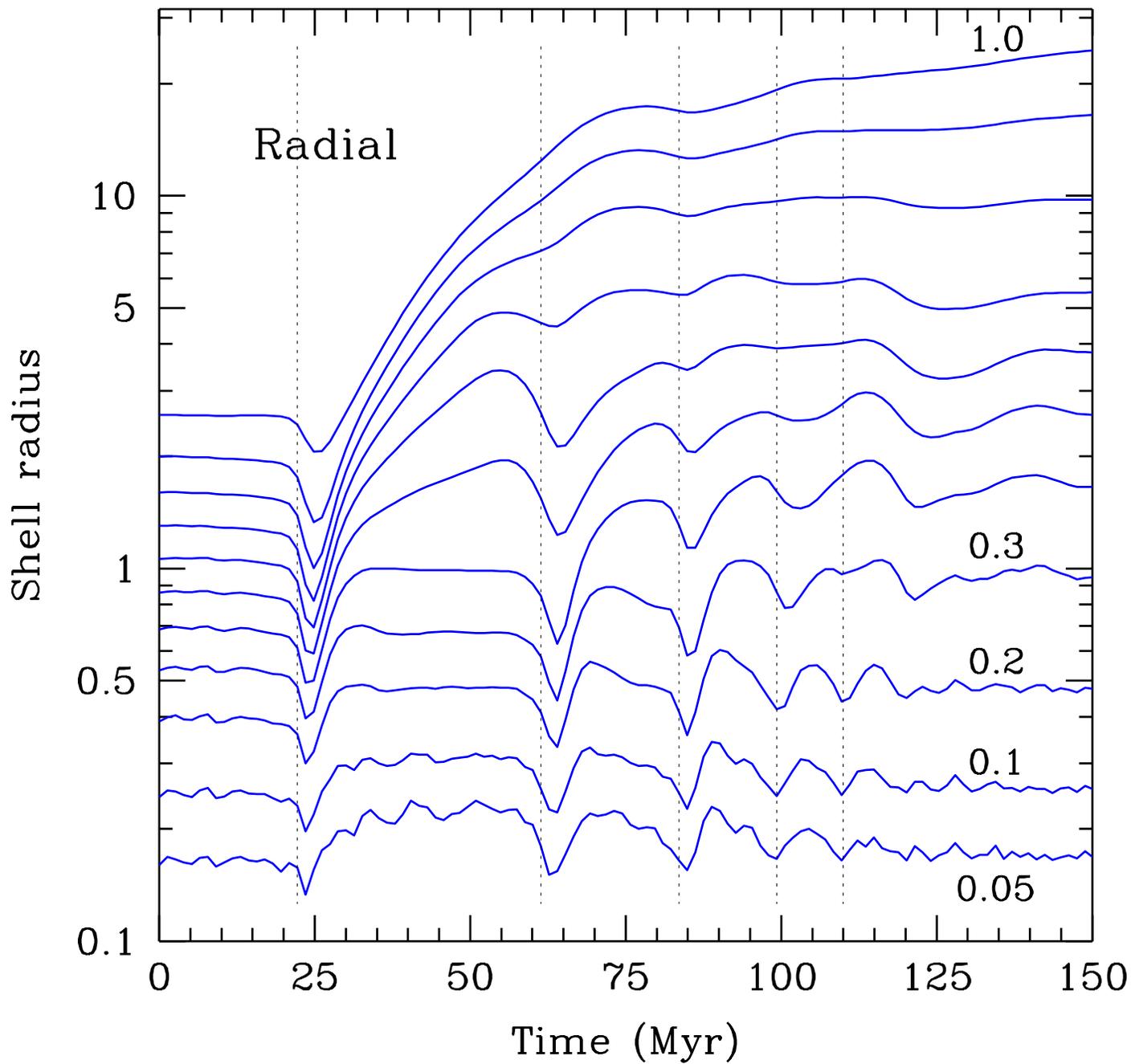
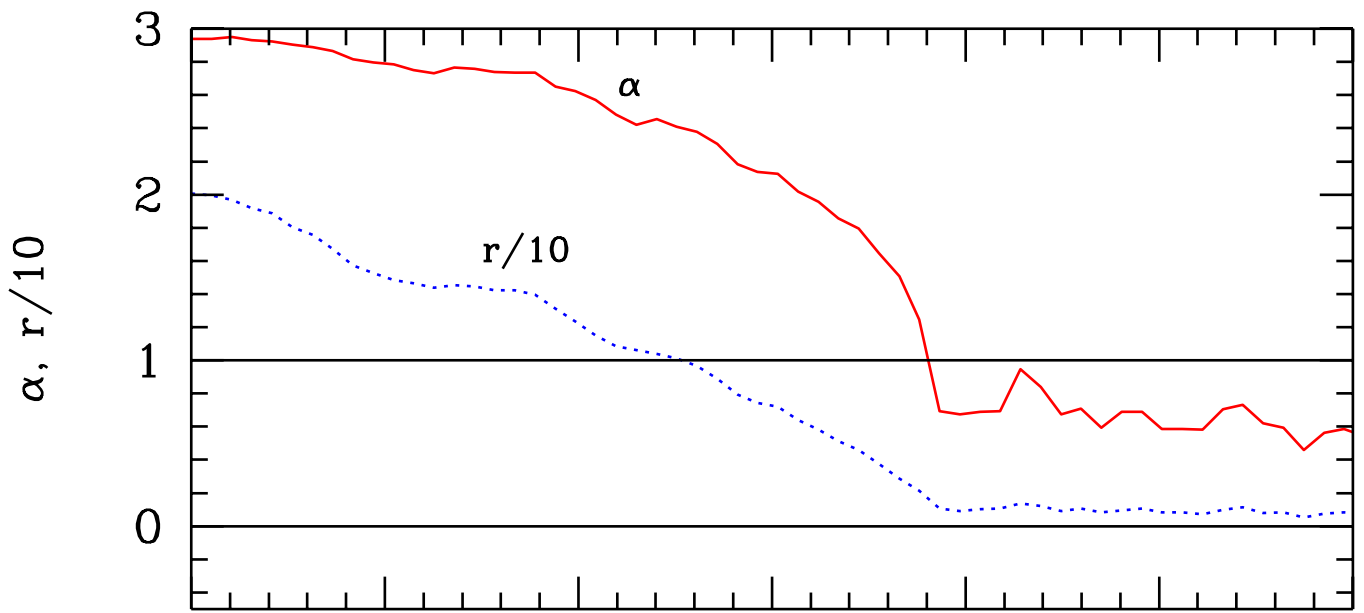
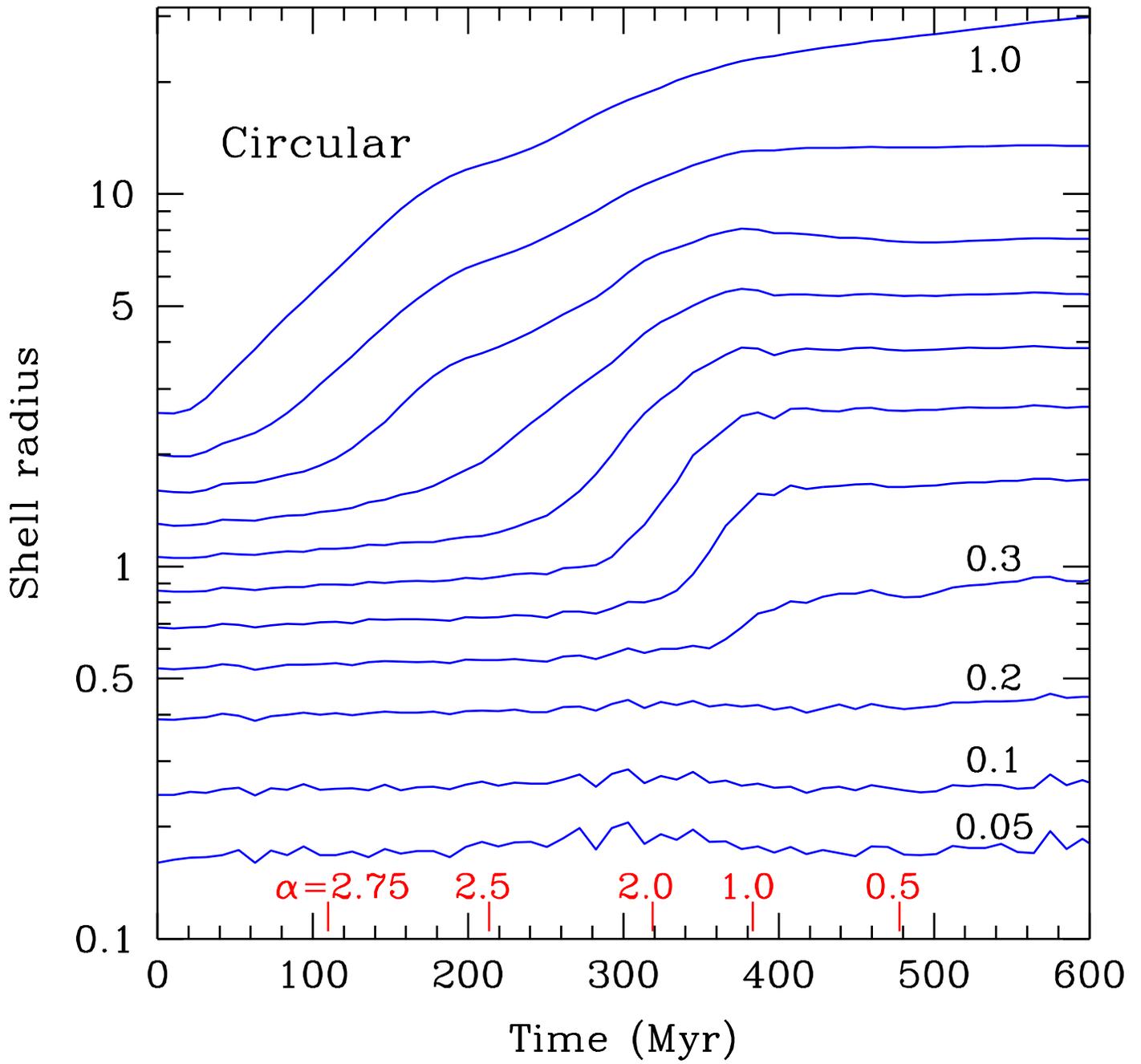
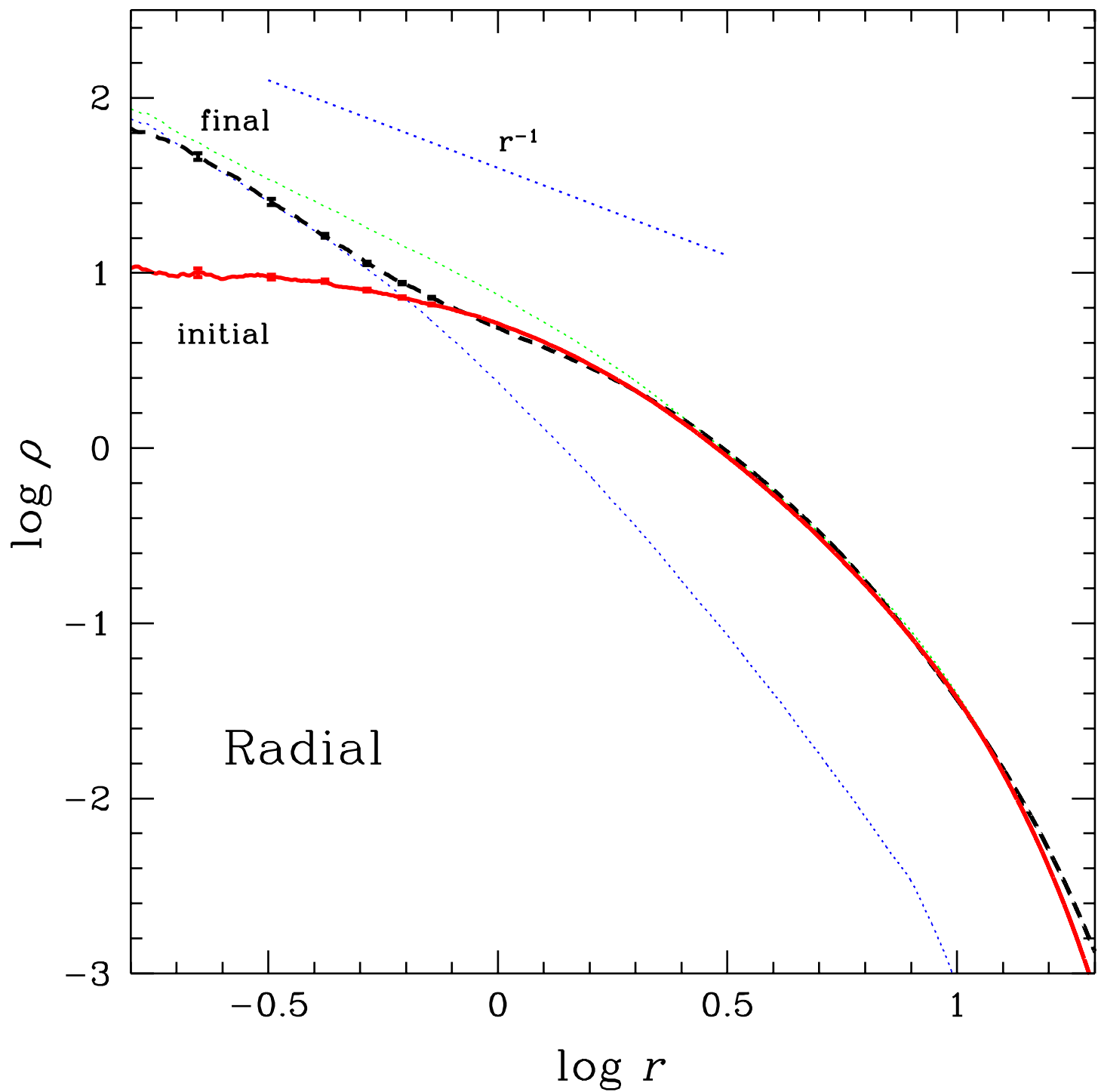
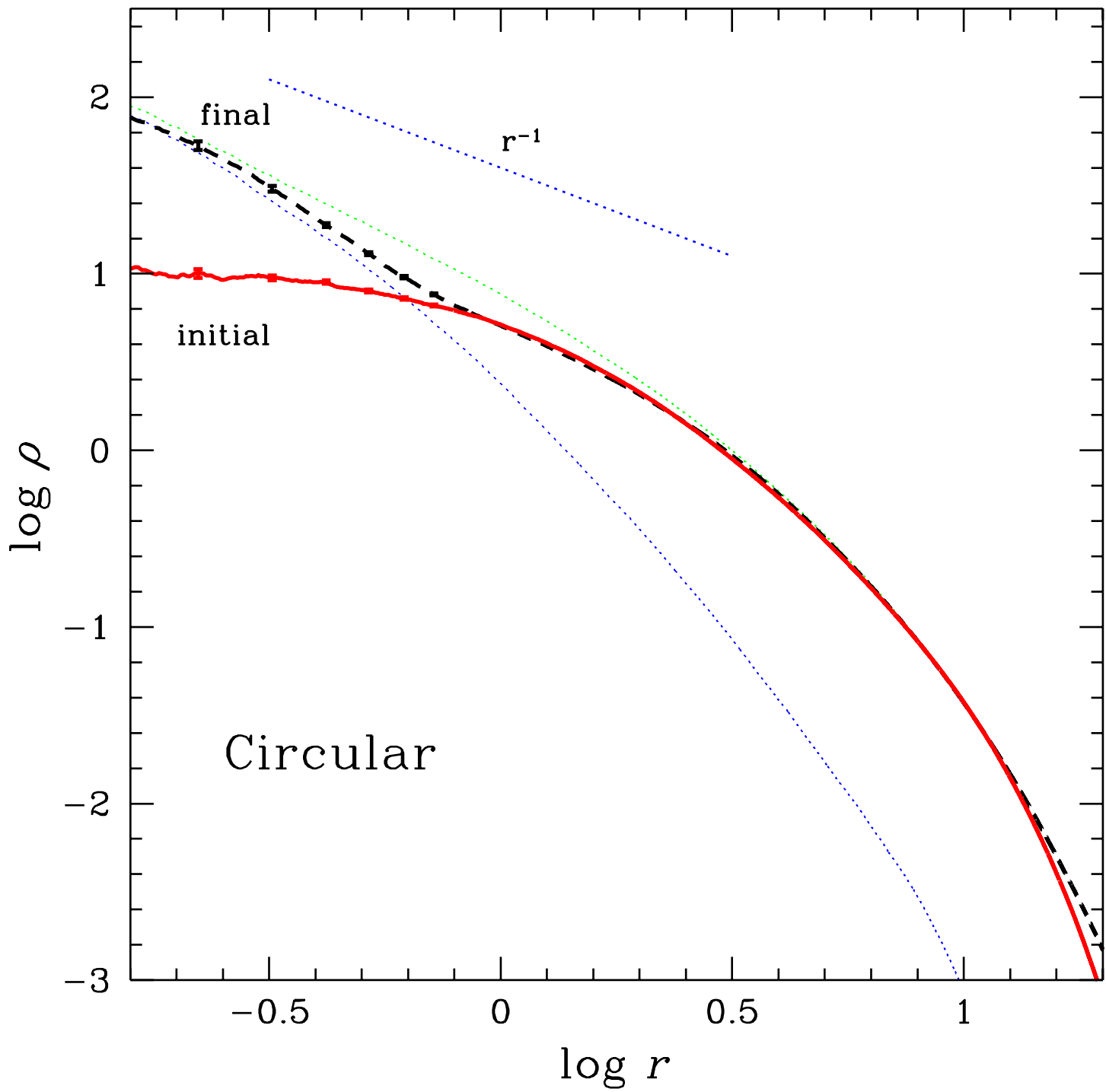
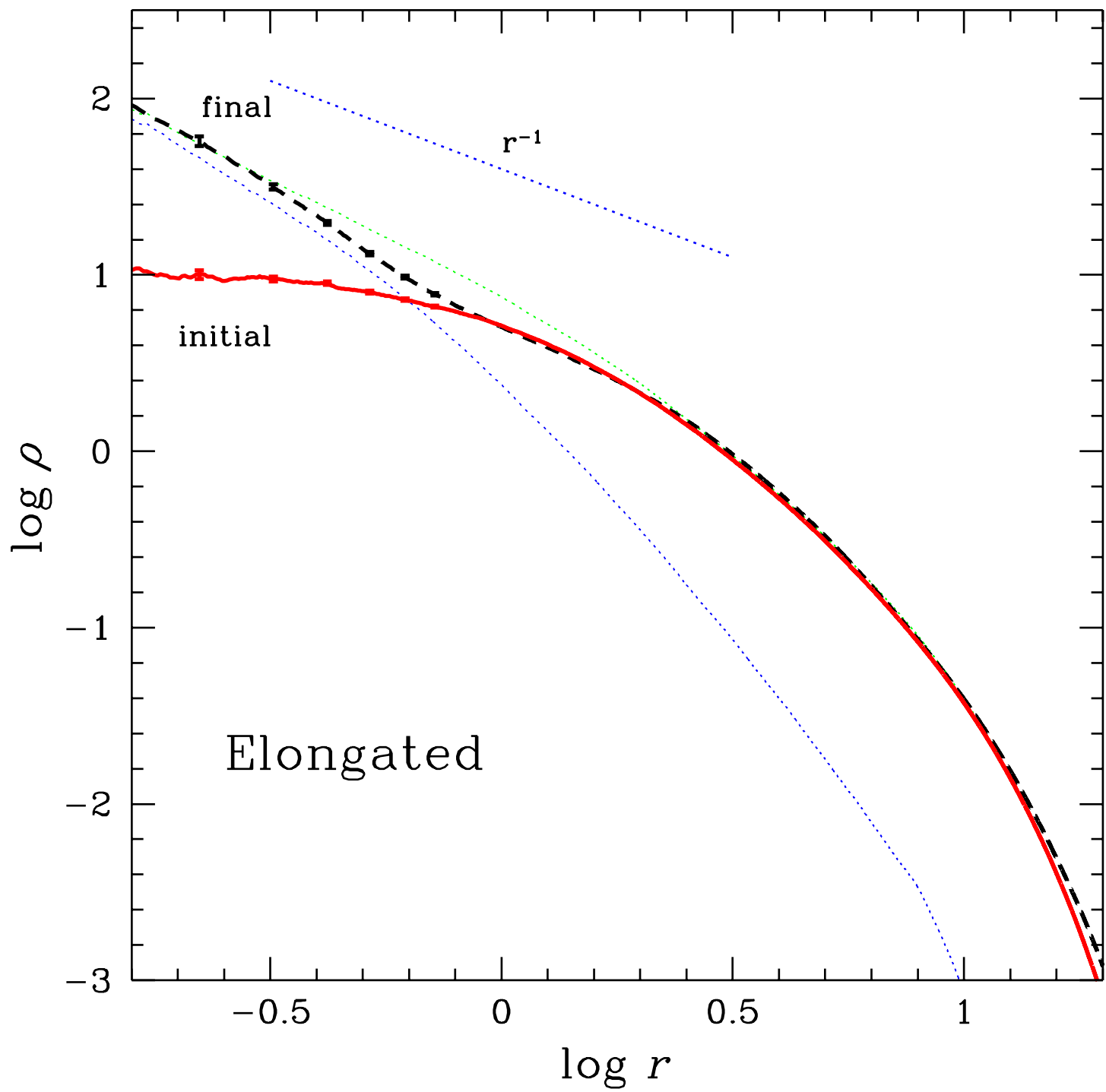
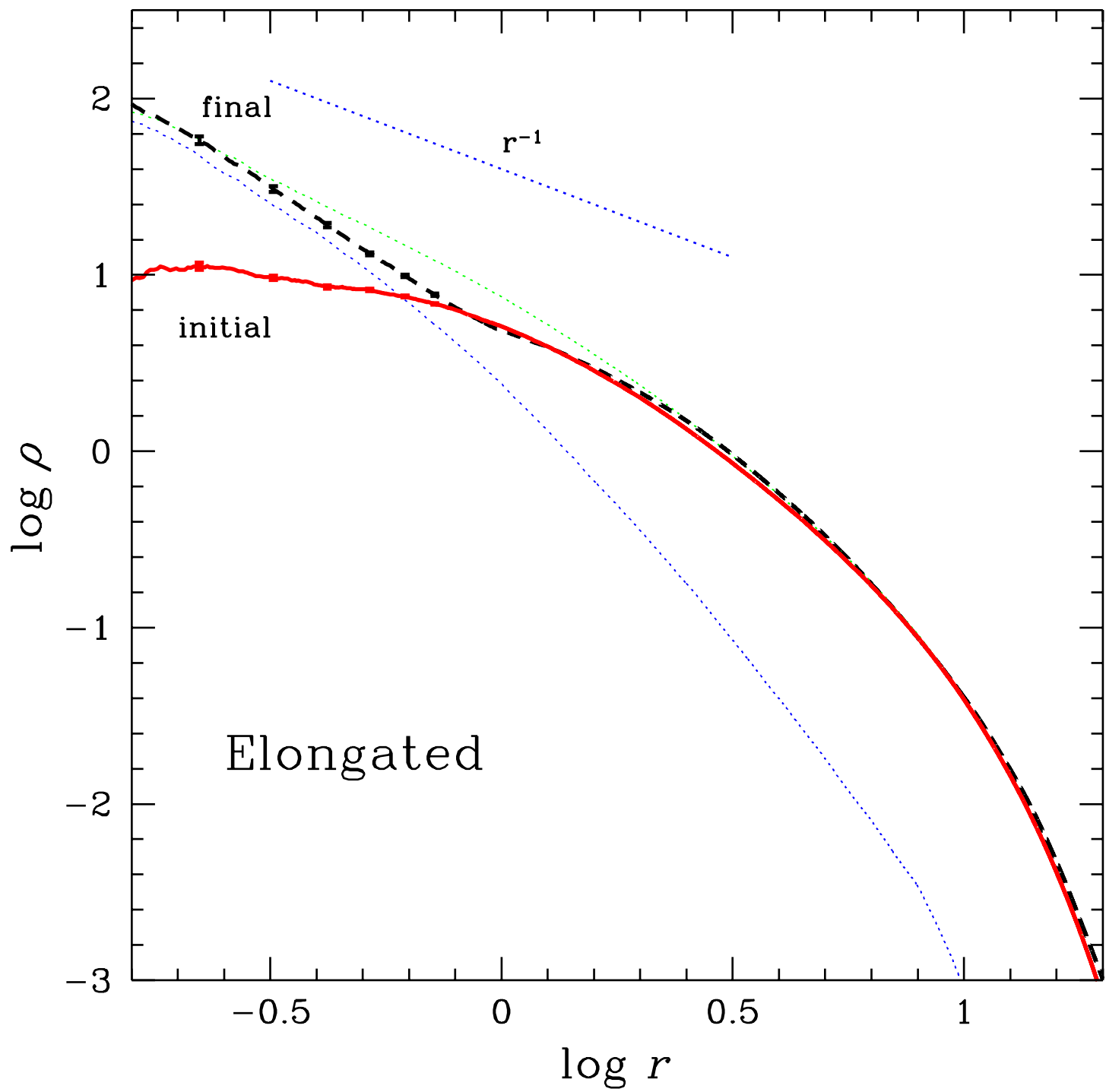
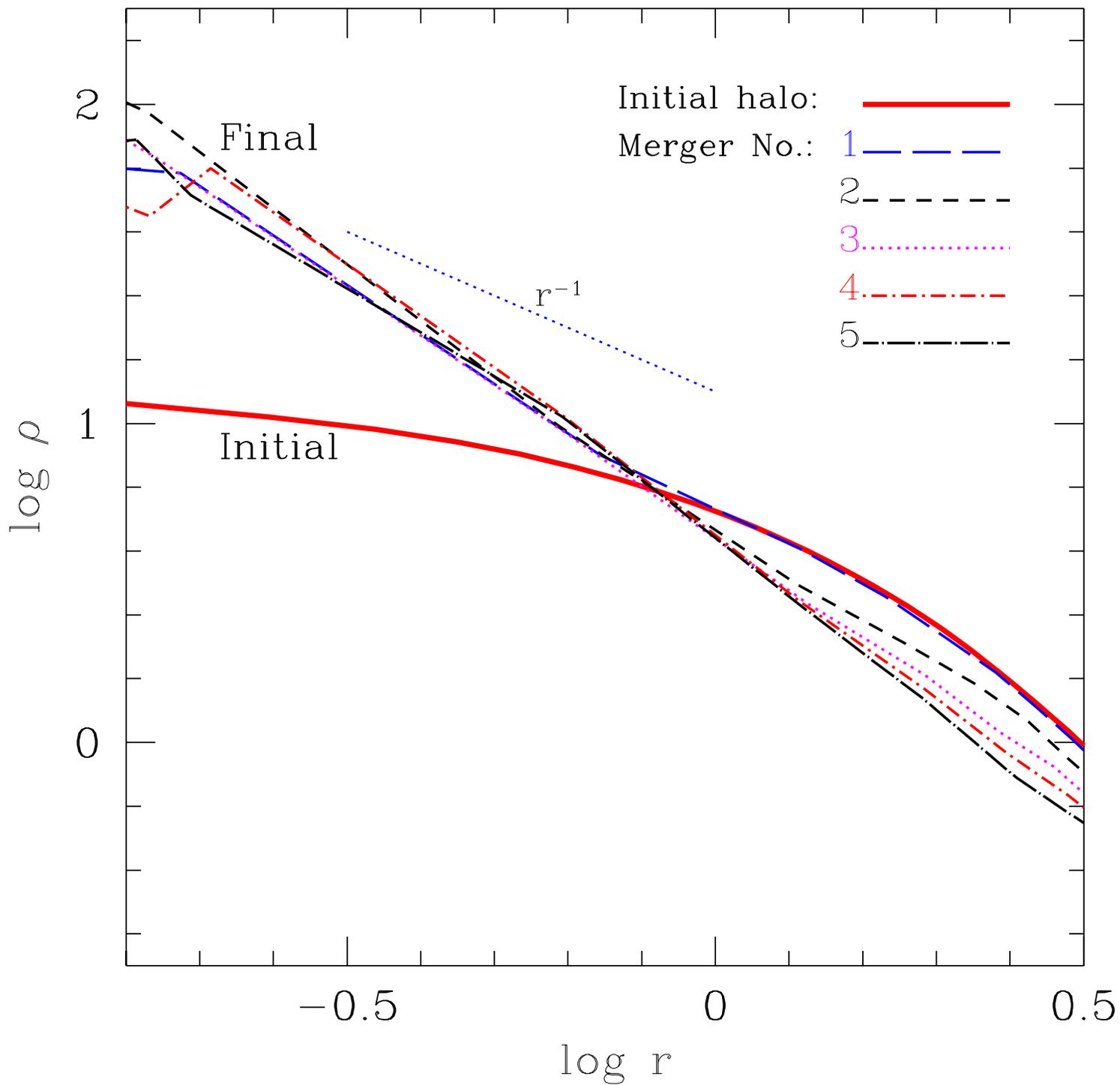
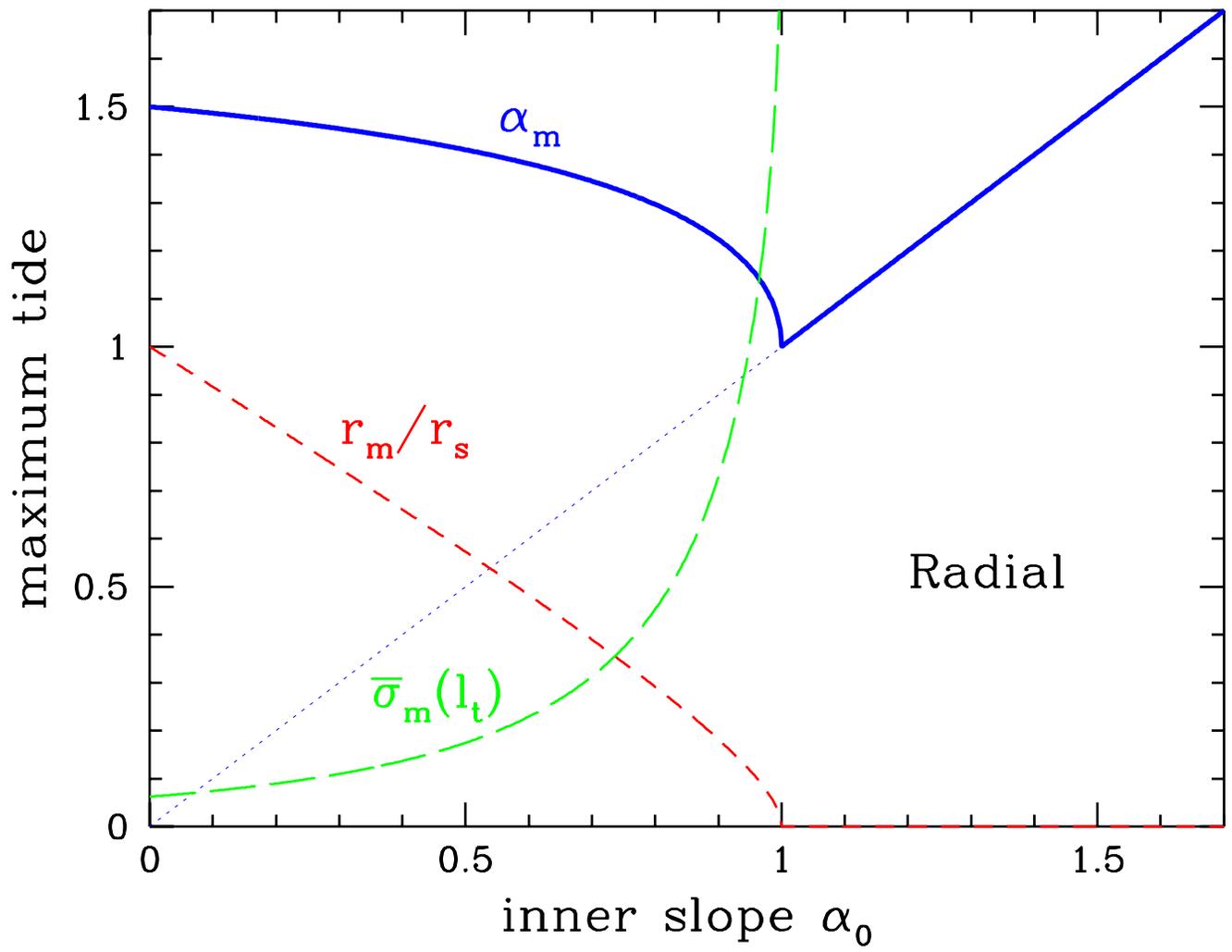
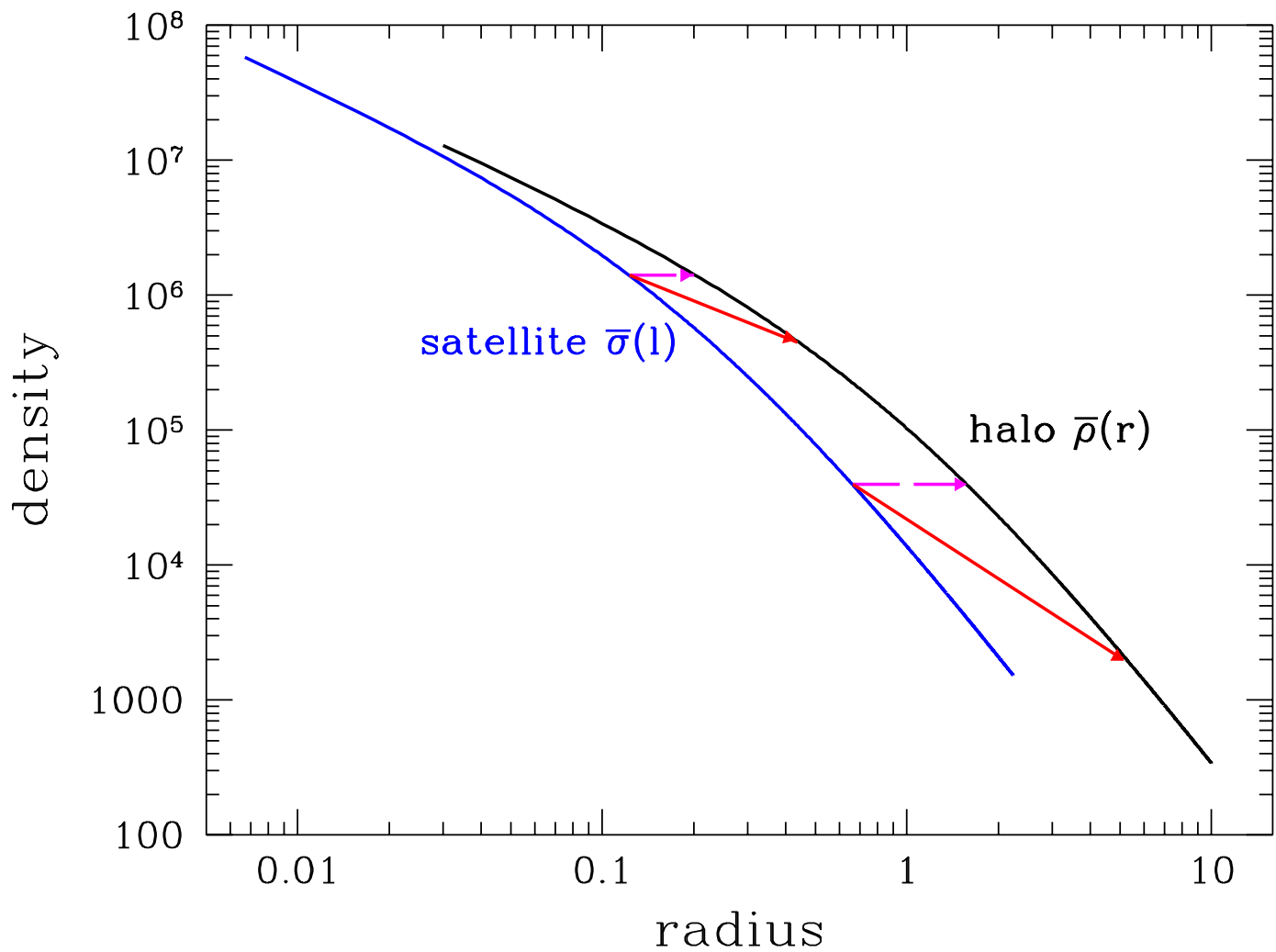
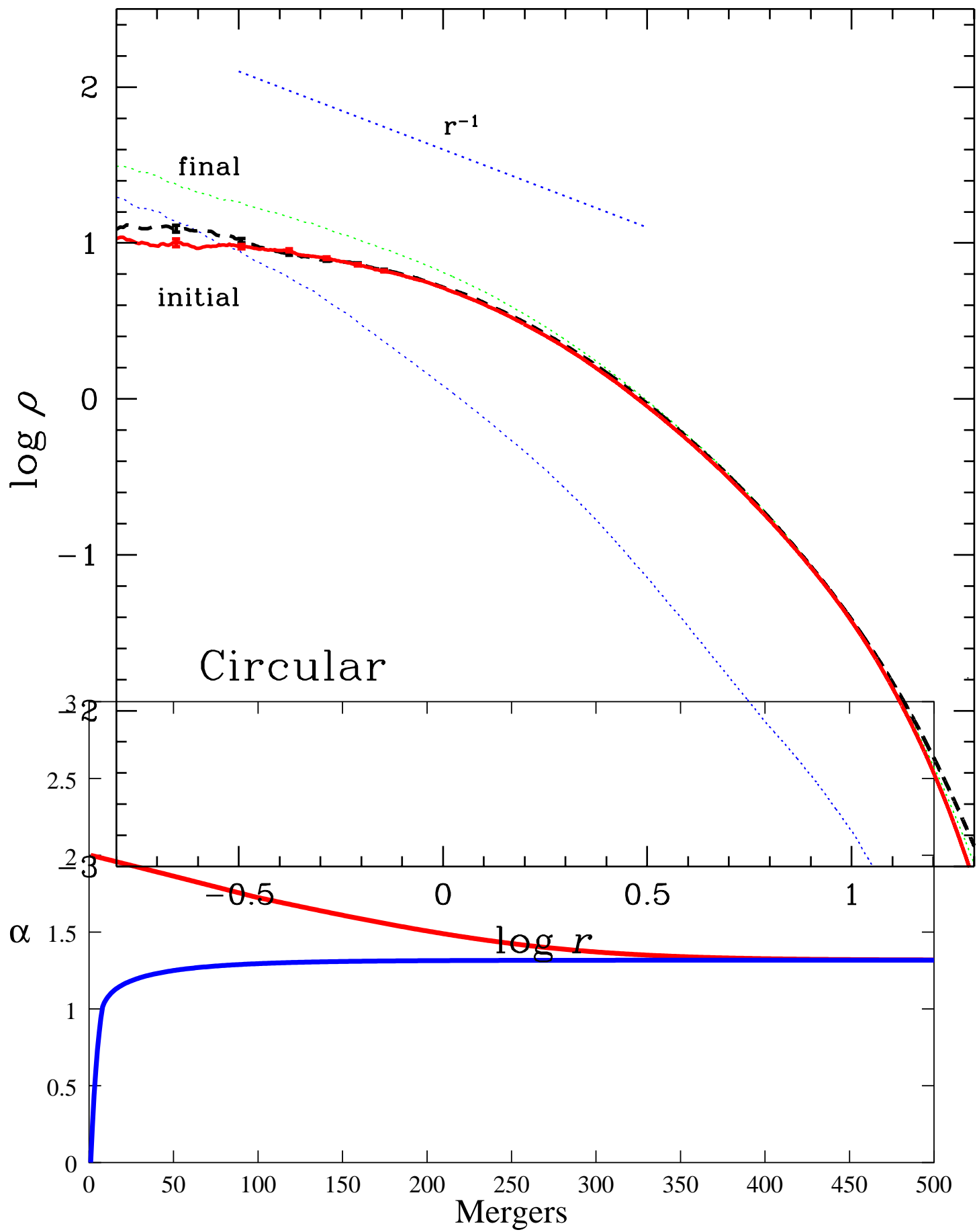
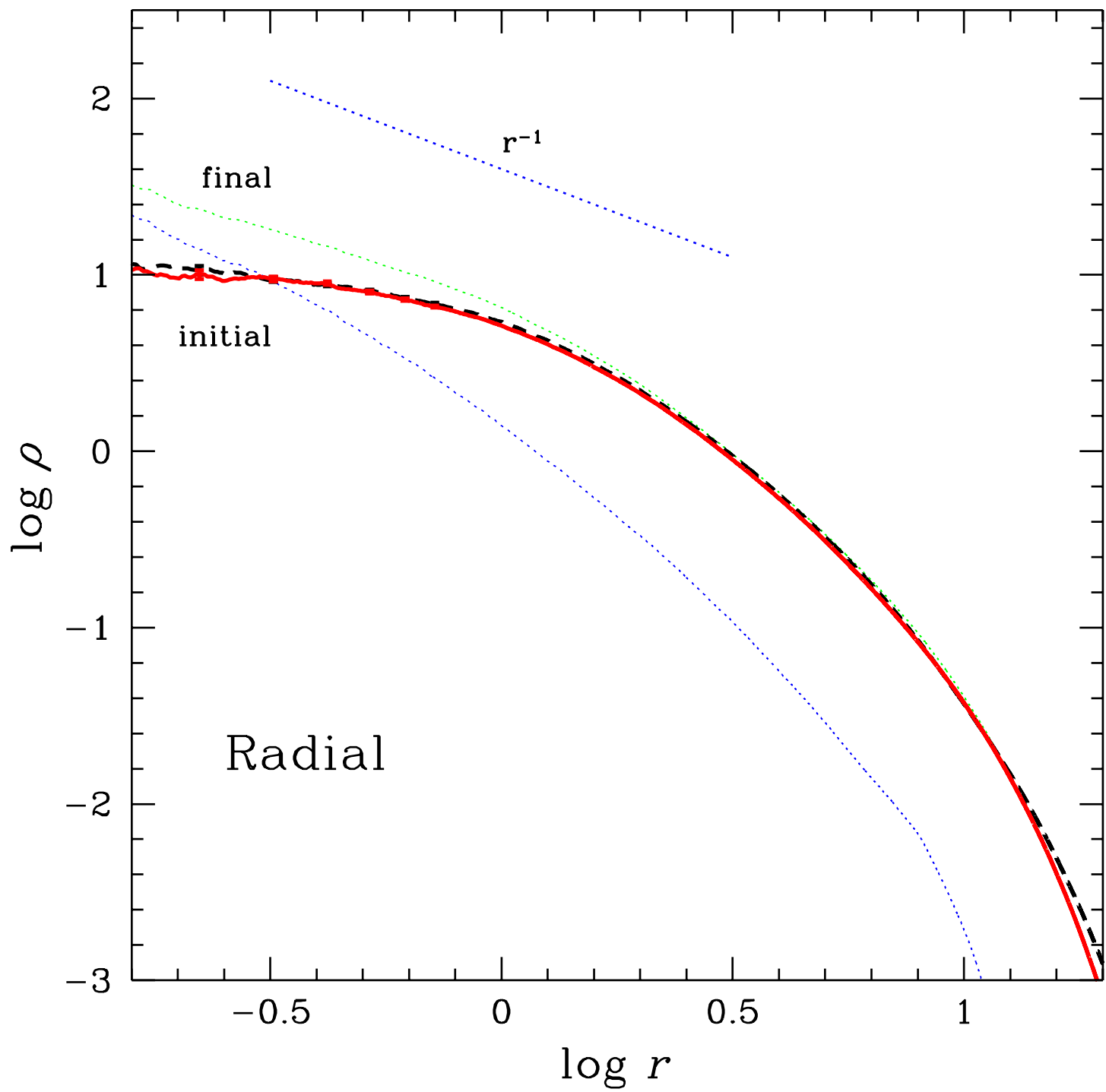
\end{document}